\documentclass[gmd,manuscript]{copernicus}

\usepackage{tabularx}

\usepackage[draft]{changes}
\definechangesauthor{Hui}
\definechangesauthor{VEL}
\definechangesauthor{PJR}
\definechangesauthor{SZ}
\definechangesauthor{XZ}
\definechangesauthor{KZ}
\definechangesauthor{WL}

\nolinenumbers

\begin{document}

\title{
CondiDiag1.0:
A flexible online diagnostic tool for conditional sampling and budget analysis 
in the E3SM atmosphere model (EAM)}


\Author[1]{Hui}{Wan}
\Author[1]{Kai}{Zhang}
\Author[1]{Philip J.}{Rasch}
\Author[1,2]{Vincent E.}{Larson}
\Author[3]{Xubin}{Zeng}
\Author[1]{Shixuan}{Zhang}
\Author[4]{Ross}{Dixon}

\affil[1]{Atmospheric Sciences and Global Change Division, Pacific Northwest National Laboratory, Richland, Washington, USA}
\affil[2]{Department of Mathematical Sciences, University of Wisconsin--Milwaukee, Milwaukee, Wisconsin, USA}
\affil[3]{Department of Hydrology and Atmospheric Sciences, University of Arizona, Tucson, Arizona, USA}
\affil[4]{Department of Earth and Atmospheric Sciences, University of Nebraska--Lincoln, Lincoln, Nebraska, USA}



\correspondence{Hui Wan (Hui.Wan@pnnl.gov)}

\runningtitle{Online conditional and budget diagnostics}

\runningauthor{Wan et al.}

\received{}
\pubdiscuss{} 
\revised{}
\accepted{}
\published{}


\firstpage{1}

\maketitle

\begin{abstract}
Numerical models used in weather and climate prediction take into account 
a comprehensive set of atmospheric processes such as the 
resolved and unresolved fluid dynamics, radiative transfer, 
cloud and aerosol life cycles, and mass or energy exchanges with 
the Earth's surface. 
In order to identify model deficiencies and improve predictive skills, 
it is important to obtain process-level understanding of the 
interactions between different processes.
Conditional sampling and budget analysis are powerful tools 
for process-oriented model evaluation, but they often require 
tedious ad hoc coding and large amounts of instantaneous model output,
resulting in inefficient use of human and computing resources. 
This paper presents an online diagnostic tool that addresses 
this challenge by monitoring model variables in a generic manner 
as they evolve within the time integration cycle.

The tool is convenient to use. 
It allows users to select sampling conditions and 
specify monitored variables at run time.
Both the evolving values of the model variables and their increments 
caused by different atmospheric processes can be monitored and archived.
Online calculation of vertical integrals is also supported.
Multiple sampling conditions can be monitored in a single simulation
in combination with unconditional sampling.
The paper explains in detail the design and implementation of the tool in 
the Energy Exascale Earth System Model (E3SM) version 1.
The usage is demonstrated through three examples: 
a global budget analysis of dust aerosol mass concentration, 
a composite analysis of sea salt emission and its dependency 
on surface wind speed,
and a conditionally sampled relative humidity budget.
The tool is expected to be easily portable to 
closely related atmospheric models that use the same or 
similar data structures
and time integration methods.
\end{abstract}


\introduction  

Atmospheric general circulation models (AGCMs) 
used in climate research and weather prediction are
simplified mathematical representations of the
complex physical and chemical processes 
driving the evolution of the Earth's atmosphere.
Despite the necessity of simplification due to 
limited computing resources,
it is highly desirable that, to the extent possible and practical,
models should be based on first principles and 
robust quantitative relationships in
atmospheric physics and chemistry,
so that the same models can reliably provide good accuracy
under historically observed atmospheric conditions
as well as in the climate of the future.
Many tools have been used for assessing the behavior and fidelity of 
the processes represented in atmospheric models. Among those,
budget analyses are a useful method for quantifying relationships 
between atmospheric processes, and
composite analyses are useful for revealing the characteristics 
of atmospheric phenomena in specific situations.
Both methods have been widely used in process-oriented model evaluation
to help identify model deficiencies and improve predictive skills. 
Carrying out such analyses, however,
often requires tedious ad hoc coding.

Consider, for example, a model evaluation study aiming at understanding the role of 
various processes in influencing the simulated atmospheric water cycle,
which involves specific humidity $q_v$ as a prognostic variable of the AGCM.
The typical way to obtain a budget of $q_v$
is to review the model source code, 
manually add extra lines of code and variables into subroutines representing parameterizations and the dynamical core
to save the rate of change (i.e., tendency) of $q_v$ 
caused by each process of interest,
and then archive those tendencies in model output.
Since modern AGCMs are sophisticated, a complete budget analysis with the finest
granularity will likely involve a  number of tendency terms. If a researcher
wishes to obtain several different views of the $q_v$ budget with different
levels of granularity (e.g., considering all stratiform cloud processes
as a single $q_v$ tendency term in one budget but breaking it down to evaluating
condensation/evaporation and rain-formation processes separately in a second
view), then the tendencies of coarser granularity will either need to be computed
from the fine-grained terms during post-processing or be calculated online and saved in additional model variables. 
Modern AGCMs often include multiple water species as prognostic variables and tens to hundreds more variables representing aerosol and gas species. Some models also include diagnostic variables 
such as isotopes and tagged water or aerosol species originating from 
different geographical regions 
\citep[e.g.,][]{WangH_2014_BC_tagging,RZhang_2015_BC_tagging,HansiSingh_2016_JAMES_water_tracer,Bailey_2019_GRL_moist_isentroic_framework_water_isotopes,WangH_2020_water_tagging_and_sea_ice_anomalies}. 
The lines of code and additional variables that are needed to monitor, assess, and diagnose tendency terms  can quickly add up to a huge number, increasing code complexity, computational overhead, and the potential for bugs both in the code and during post-processing.
An AGCM also often contains many diagnostic variables that are
needed in the equations of a parameterization.  
For example, the relative humidity with respect to ice (RHI) is often used 
in the prediction of formation of cloud ice crystals
 (cf. Sect.~\ref{sec:use_case_RHI}).
While an AGCM might only calculate RHI
once or a few times during each time step, 
a detailed budget analysis of the terms influencing RHI can provide useful insights
into the atmospheric processes that contribute to
or compete with ice cloud formation.
These types of diagnostic variables appear frequently in AGCMs, and supporting budget analyses for them would require 
inserting many new model variables and output, 
which often leads to a
dilemma in source code management:
that if a user throws away the ad hoc coding after 
their study is completed,
other users interested in similar topics will 
need to reinvent the wheel or at least re-do the coding;
on the other hand, if users 
commit study-specific code to the model's central repository, 
clutter will accumulate quickly.

Similar challenges are encountered in studies involving composite analysis,
the essence of which is to define a criterion, 
conditionally sample some model variables,
and then analyze the stratified data
to look for relationships occurring under specific conditions.
Conditional sampling in AGCM simulations is often carried out 
by first archiving a large amount of instantaneous model fields 
at a sufficiently high frequency, and then using post-processing 
to produce the conditionally sampled composite \citep[see, e.g.,][]{ghan:2016,gryspeerdt:2020}.  
This not only can lead to inefficient use of computing time 
(due to I/O bottleneck) but also creates challenges in data storage and transfer. 
Occasionally, conditional sampling is carried out online
(i.e., during a simulation) so that only the temporal averages of model variables
meeting the sampling condition need to be archived.
With this approach, ad hoc coding is often used
for each combination of sampling condition and monitored variable,
which again results in challenges in code management.

Authors of the present paper recently started an effort
to identify and address numerical artifacts in
time integration related to physics parameterizations 
and process coupling in version 1 of the E3SM atmosphere model
\citep[EAMv1,][]{Rasch_et_al:2019,Xie_et_al:2018}.
The study of \citet{Wan_2021_GMD_time_step_sensitivity}
and its follow-up investigations have involved monitoring 
not only EAM's prognostic variables but also  
non-standard output fields such as various measures of 
supersaturation and atmospheric instability. Those investigations constantly require the use of composite and budget analyses,
motivating our development of a new, general, and user-friendly online diagnostic tool to facilitate the investigations.
This paper presents the first version of the new tool, which we refer to as CondiDiag1.0.

Assuming the physical quantities to be monitored 
already exist in EAM,  
configuring a simulation to activate CondiDiag will normally 
require only setting a small number of switches in the model's input file (currently using Fortran namelist, cf.~Sect.~\ref{sec:namelist}).
A minimal amount of special-purpose code might be required from
the user if existing model variables need to be monitored 
at new locations in the model's time loop, or
if the variables exist within a parameterization or the dynamical core
but need to be made available in the data structures accessible 
by our tool; the coding required in such cases will be simple.
To facilitate budget analyses,
the tool provides the flexibility to monitor and archive
both the evolving values of model variables and their increments 
caused by different atmospheric processes.
Vertical integrals can be calculated online .
Multiple sampling conditions can be used in the same simulation. 
Unconditional sampling and  mixtures of conditional and unconditional sampling 
are also supported.

The new tool has been designed for and implemented in 
EAMv1 and ported to a development version of EAMv2.
It should also be straightforward to port it to 
EAMv1's recent predecessors,
e.g., the Community Atmosphere Model versions 5 and 4 
\citep[CAM5 and CAM4,][]{neale:2012,neale:2010}, 
as these models use the same Fortran derived data types
for organizing information passed through the 
physics parameterizations suite. 
Examples of such Fortran data types include
the ``physics state'', ``physics buffer'', 
atmosphere ``import'' and ``export'' variables.
It is also possible to revise our tool for implementation in other models, 
as the underlying design concepts are generalizable.

The remainder of the paper is organized as follows:
Section~\ref{sec:EAM} introduces 
features of EAMv1's time integration schemes and output capability that our tool makes use of.
Section~\ref{sec:design} introduces the key concepts
and basic design of our tool.
Section~\ref{sec:implementation} describes the implementation of CondiDiag 
in EAMv1 and Sect.~\ref{sec:users_guide} 
provides a brief user's guide.
Section~\ref{sec:examples} presents three concrete examples
to further demonstrate the usage of the tool: 
a global budget analysis of dust aerosol mass concentration, 
a composite analysis of sea salt emission and its dependency on surface wind speed,
and a conditionally sampled relative humidity budget.
Section~\ref{sec:conclusions} summarizes the paper 
and points out possible future improvements and extensions of the tool.


\section{Host model features}
\label{sec:EAM}

Here, ``host model" refers to the AGCM in which our new tool is
embedded, in this case EAMv1.
We summarize EAM's choice of method for coupling atmospheric processes 
in Sect.~\ref{sec:sequential_splitting} and 
briefly describe how model variables are archived on output files
in Sect.~\ref{sec:history_output_from_EAMv1}. 
These features of the host mode are used by our tool.

\subsection{Sequential process coupling}
\label{sec:sequential_splitting}

\begin{figure*}[t]
\includegraphics[height=15cm]{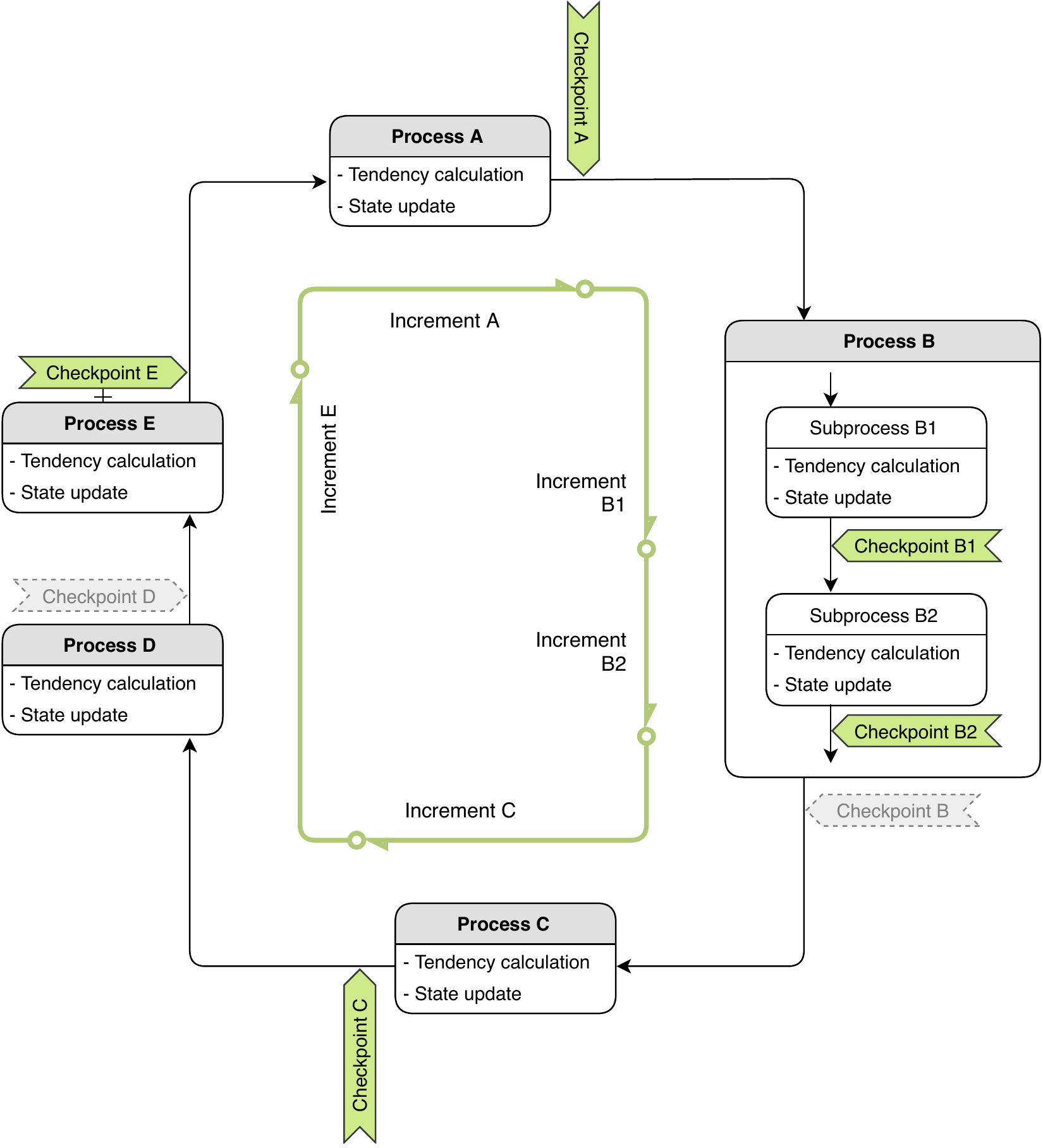}
\caption{A schematic showing 
a time step of model simulation involving five hypothetical atmospheric processes, 
A to E, either resolved or unresolved by the model's computational mesh,
that are numerically coupled using isolated sequential splitting
(cf. Sect.~\ref{sec:sequential_splitting}).
Also shown are various locations (referred to as checkpoints,
cf.~Sect.~\ref{sec:checkpoint_concept}) within a time step that are introduced to facilitate diagnostics using the new tool.
When the tool is used in a simulation, some checkpoints 
are activated (i.e., selected by the user and indicated 
in green here in, and others
are inactive and indicated in gray.
No information is monitored at inactive checkpoints. 
The green lines with a circle on one end and an arrowhead 
on the other end depict how increments of model variables 
are defined. Further details can be found in Sections~\ref{sec:sequential_splitting} and \ref{sec:checkpoint_concept}.
}
\label{fig:checkpoints}
\end{figure*}

EAMv1 solves a set of integral-differential equations to simulate
the spatial variation and temporal evolution of the state of 
the atmosphere. Distinct physical and chemical processes 
are represented by different model components, e.g., 
the dynamical core that describes the large-scale fluid dynamics
and tracer transport processes resolved by the model's computational
grid, and various parameterizations that describe smaller-scale 
fluid processes (e.g., turbulence and convection) and non-fluid phenomena (e.g., radiative transfer, 
cloud and aerosol microphysics). 

For time integration, the primary method 
used in EAMv1 for component coupling is 
a method we refer to as isolated sequential splitting
(Figure~\ref{fig:checkpoints}).
In this method, a model component produces an estimate
of the rate of change of the atmospheric state by considering
a single or a set of closely related physical or chemical processes in isolation 
(i.e., ignoring all other processes represented by other model components). 
The estimated rate of change, often referred to as a tendency, 
is used to update the atmospheric state, and then the updated
state is passed to the next model component. 
Since EAMv1 contains many components, 
the atmospheric state is updated multiple times within one full time step. 
Here a full time step is defined as the smallest time integration cycle 
in which the effects of all physical processes considered in a simulation
have been used to update the model state 
at least once in advancing the solution in time. 
This full time step is often loosely referred to 
as the ``physics time step'' in EAMv1 and its predecessors.
In a discussion on time stepping and sub-cycling in EAMv1, 
\citet{Wan_2021_GMD_time_step_sensitivity} referred to the full time steps 
as the ``main process-coupling time steps'' and denoted their length by $\Delta t_{\rm CPLmain}$.
The same notation is used in this paper for consistency and clarity.
The so-called low-resolution configuration of EAMv1 (with 1~degree horizontal grid spacing) uses $\Delta t_{\rm CPLmain} =$~30~min by default.
Fig.~\ref{fig:checkpoints} provides a schematic showing a full time step 
consisting of 5 hypothetical processes label as A to E.

A model component in EAMv1 might contain sub-components that are also connected using the isolated sequential splitting method, like process B depicted in Fig.~\ref{fig:checkpoints}. 
An example in EAMv1 is deep convection, 
which consists of the parameterization by \citet{Zhang_McFarlane:1995} 
that describes impact of convective activities on temperature and humidity
and a parameterization of the convective momentum transport from \citet{Richter_Rasch:2008}. These convection-related atmospheric processes are sequentially split within the deep convection parameterization.

Another situation that can also be depicted by the hypothetical
process B in Fig.~\ref{fig:checkpoints} is sub-cycling.
For example, in EAMv1, the parameterizations of turbulence, shallow convection, and stratiform cloud macrophysics and microphysics are sub-cycled 6 times within each 30~min full time step. In this case, each sub-cycle can be viewed as a sub-process depicted in Fig.~\ref{fig:checkpoints} (i.e., sub-cycle 1 corresponds to process B1, sub-cycle 2 corresponds to process B2, etc.).

\subsection{History output}
\label{sec:history_output_from_EAMv1}

EAMv1 inherited from its predecessors a flexible mechanism for 
handling model output \citep[see, e.g.,][]{CAM5.1_user_guide}.
The data files that contain the temporal and spatial distribution
of model-simulated physical quantities are called history files.
The model can write multiple series of history files with 
different write frequencies; these series are referred to as 
history tapes in the source code.
Different history tapes can contain different output 
variables (fields).
Whether the values written out should be instantaneous, 
time-averaged, maximum or minimum during the output time window
can be specified for each tape on a field-by-field basis.

The software infrastructure for history output  uses internal data types
and functions that handle the storage of fields to be 
written out and perform the calculation of required statistics 
(e.g., time averages). 
Typically, 
researchers focusing on physical or computational aspects
of the model do not need to care about the internal working
of this software infrastructure.
Rather, they use a subroutine named \texttt{outfld} 
to transfer the values of a model variable to the infrastructure.
To provide a context for some descriptions in later sections,
we note that while a model variable can change its value multiple times in a time step of $\Delta t_{\rm CPLmain}$, the value being recorded 
for output is the snapshot made when the \texttt{outfld}
subroutine is called. 
The location in the time integration cycle at which 
the \texttt{outfld} subroutine is called can differ
from model variable to variable.

\section{Nomenclature and design concepts for CondiDiag}
\label{sec:design}

We now introduce the key concepts and design features of 
the new tool. The description in this section is kept general, 
only referring to EAM when necessary,
as the methodology can be applied to
or revised for other AGCMs.
Details of the implementation in EAMv1 are
provided in Sect.~\ref{sec:implementation}.

\subsection{Checkpoints, field values, and increments}
\label{sec:checkpoint_concept}

In order to discuss the implementation of our tool in the context of 
the sequential process splitting described 
in Sect.~\ref{sec:sequential_splitting}, we introduce
the following nomenclature:
\begin{itemize}
\item A {\it checkpoint} is a location in the time integration cycle where 
a snapshot of a model variable can be obtained (cf.~Fig.~\ref{fig:checkpoints}). 
At a checkpoint, the value of a model variable can be retrieved from
data structures in the standard EAMv1.
Additional quantities can be computed from available variables.
Those retrieved or computed variables at that checkpoint can be
saved in the data structure of our tool and
transferred to the output-handling infrastructure 
of the standard EAM (cf. Sect.~\ref{sec:history_output_from_EAMv1}).
If a process is sub-cycled with respect to $\Delta t_{\rm CPLmain}$, 
then the end of each sub-cycle is considered to be
a different checkpoint.

\item The value of a model variable at a checkpoint is 
referred to as a {\it field value}. For example, the 
air temperature after process A in Fig.~\ref{fig:checkpoints}
is referred to as the field value of temperature at checkpoint A.

\item Checkpoints available in the model are active or inactive in a simulation:
Active checkpoints are selected by user at run time. 
No information is retrieved, calculated, or archived 
by our tool at inactive checkpoints.
This flexibility allows a user to focus only on the checkpoints 
relevant to their specific study;
it also saves memory and disk space, 
as inactive checkpoints will not consume memory 
or produce information in the model's output files. 

\item The difference between values of the same model variable
at two different checkpoints is referred to as an 
{\it increment}. 
Since there can be inactive checkpoints, an 
increment calculated by our tool is the difference
between the field value at the current checkpoint 
and the field value at the previous active checkpoint. 
For example, in Fig.~\ref{fig:checkpoints}, increment E is the
difference between checkpoints E and C, 
with the inactive checkpoint D ignored. 
\end{itemize}

\subsection{Composite analysis}
\label{sec:composite_analysis}

For a composite analysis, our tool expects the user to 
specify one or more conditional sampling criteria via
run time input (e.g., namelist parameters).
The handling of multiple conditions 
is described later in Sect.~\ref{sec:multiple_conditions}.
Here we first explain the handling of a single sampling condition.

During each time integration cycle of length $\Delta t_{\rm CPLmain}$, 
values of user-selected variables at active checkpoints 
are obtained and copied to a data structure internal to our tool.
Increments and vertical integrals are calculated if requested.
The sampling condition is evaluated at each grid cell in the global domain.
Depending on whether the condition is met,
the copy (including increments and integrals)
is assigned either the model-computed values or a fill value,
resulting in a conditionally sampled copy.
This sampled copy, 
together with information about the sampling condition,
is then transferred to the output handling infrastructure.
In the next time step, the sampling condition is re-evaluated and 
selected model variables re-sampled. The details are explained below.

\subsubsection{Defining a condition}
\label{sec:defining_a_condition}

A key element of a sampling strategy is the 
atmospheric condition to be used to categorize data.
Necessary elements in the definition of a condition include
(1) a {\it metric} (which can be any 2D or 3D field,
e.g., air temperature or surface pressure), 
(2) a {\it threshold} (which is a number, 
e.g., -40~$^\circ$C or 500~hPa), 
and (3) a {\it comparison type} (e.g., smaller than or equal to). 
In our tool, a metric can be any prognostic or
diagnostic variables in the host model or 
a quantity that can be diagnosed from existing variables. 
Currently supported comparison types include (i) $<$, (ii) $\leqslant$, (iii) $>$, (iv) $\geqslant$, 
and (v) equal to within a tolerance. 
Type (v) can be used to select values within a range.
For example, choosing a threshold of -20$^\circ$C 
and a tolerance of 20~$^\circ$C would allow the user to
sample grid cells with air temperature between -40$^\circ$C 
and 0~$^\circ$C.
The user's choices of metric, threshold, comparison type, and tolerance (if applicable)
are expected to be specified through run time input. 

Another key element of the definition of the sampling condition is
the location in the time integration cycle at which the 
sampling condition should be evaluated. 
As explained earlier in Sect.~\ref{sec:sequential_splitting},
the atmospheric state defined by the prognostic variables of 
EAM's governing equations is updated multiple times 
within one full time step of $\Delta t_{\rm CPLmain}$
due to the sequential method used for process coupling.
While some diagnostic quantities (e.g., relative humidity) 
are calculated only once or a few times per $\Delta t_{\rm CPLmain}$
when they are needed by a calculation (e.g., a parameterization),
their values that are consistent with the model's prognostic state 
effectively evolve within each time step.
To avoid ambiguity, our tool requires the user to specify 
at which checkpoint (cf.~Sect.~\ref{sec:checkpoint_concept})
a sampling condition should be evaluated.
The implementation of this aspect in EAMv1 is discussed 
in more detail in Sect.~\ref{sec:namelist_conditions}.

\subsubsection{Condition metric and field of flags}
\label{sec:metric_and_flag_field}

In this first version of our tool, 
the metric used in defining a sampling condition can be one of the following 
types of model variables:
\begin{itemize}

\item a 2D field that covers the entire horizontal domain of the model,
such as the surface pressure or total cloud cover;

\item a 3D field defined at layer mid-points or as layer averages,
e.g., air temperature, cloud fraction, or the mass mixing ratio 
of a tracer in EAMv1,

\item a 3D field defined at the interfaces between adjacent layers,
e.g, the convective mass flux predicted by the deep convection parameterization
or the net longwave radiative flux calculated by the radiation scheme in EAMv1.

\end{itemize}
For each condition metric, a flag field with the same spatial dimensions 
is defined in the data structure internal to our tool.
After a sampling condition is evaluated at a grid cell in the 2D or 3D domain, 
the flag field is assigned 
a value of 1 if the condition is met and a value of 0 otherwise.
The flag field, when averaged over time,
quantifies the frequency of occurrence of meeting the sampling condition
at each individual grid cell.
The flags at different grid cells can be averaged in space,
either over the entire 2D or 3D domain or over a subdomain,
to calculate the frequency of occurrence of the sampling condition
in the corresponding domain, but the spatial averages are expected to be
done during post-processing instead of during model integration.
A use case example involving both temporal and spatial averaging 
can be found in Sect.~\ref{sec:use_case_RHI}.

After the sampling condition is evaluated over the entire 2D or 3D domain,
the condition metric itself is sampled, meaning that 
the field of values transferred to the output-handling software
contains the model-computed values where the condition is met and 
a fill value of zero where the condition is not met.
In other words, the masking indicated by the flag field is applied to the 
condition metric as well.
Recall that the output-handling infrastructure of EAM 
supports both instantaneous and time-averaged model output.
Since EAM is a climate model, 
time-averaged output is expected to be more often used.
Our tool uses a fill value of zero for archiving the condition metric 
and the other monitored model variables 
to make sure that time steps in which the sampling condition 
is not met make zero contributions
to the time average. Later on, during post-processing,
when a time average of a condition metric is divided by 
the time average of the corresponding flag,  
we get the composite mean, i.e., the average over 
time steps when the condition is met.

\subsubsection{Monitored model variables}

Our tool allows for multiple model variables to be monitored under the 
same sampling condition. To distinguish those monitored variables
from the condition metric, the monitored variables are 
referred to as the quantities of interest (QoIs) in the remainder of this 
paper and in our code. QoIs monitored under the same condition
can have different vertical dimension sizes:
\begin{itemize}
\item When the QoI has the same dimension size as the condition metric,
the masking indicated by the flag field can be
applied in a straightforward manner. 

\item If the metric is 2D and the QoI is 3D, then the same 2D masking
is applied to all vertical layers or interfaces.

\item If the metric and the QoI are both 3D but have different numbers 
of vertical layers
(e.g., the metric is the air temperature defined at layer midpoints
while the QoI is the net longwave radiative flux defined at layer interfaces), 
then masking will be skipped, 
meaning this specific QoI will be captured for output 
as if no conditional sampling had happened.

\item If the metric is 3D and the QoI is 2D,  
then a grid cell in the 2D domain is 
selected if any layer midpoint or interface in that column is selected.
For example, to quantify the shortwave cloud radiative effect (the QoI)
in the presence of ice clouds, one can 
choose a sampling condition of non-zero ice crystal concentration.
Then, if ice crystals occur in any layer in a grid column, 
then the shortwave cloud radiative effect of that grid column
will be sampled.

\end{itemize}
Like the archiving of the condition metric
explained in Sect.~\ref{sec:metric_and_flag_field}, 
a QoI gets a fill value of zero at grid cells 
where the condition is not met, so that the composite mean 
can be derived by dividing the time-averaged QoI by the 
time-averaged flag field.

\begin{figure*}[t]
\includegraphics[height=15cm]{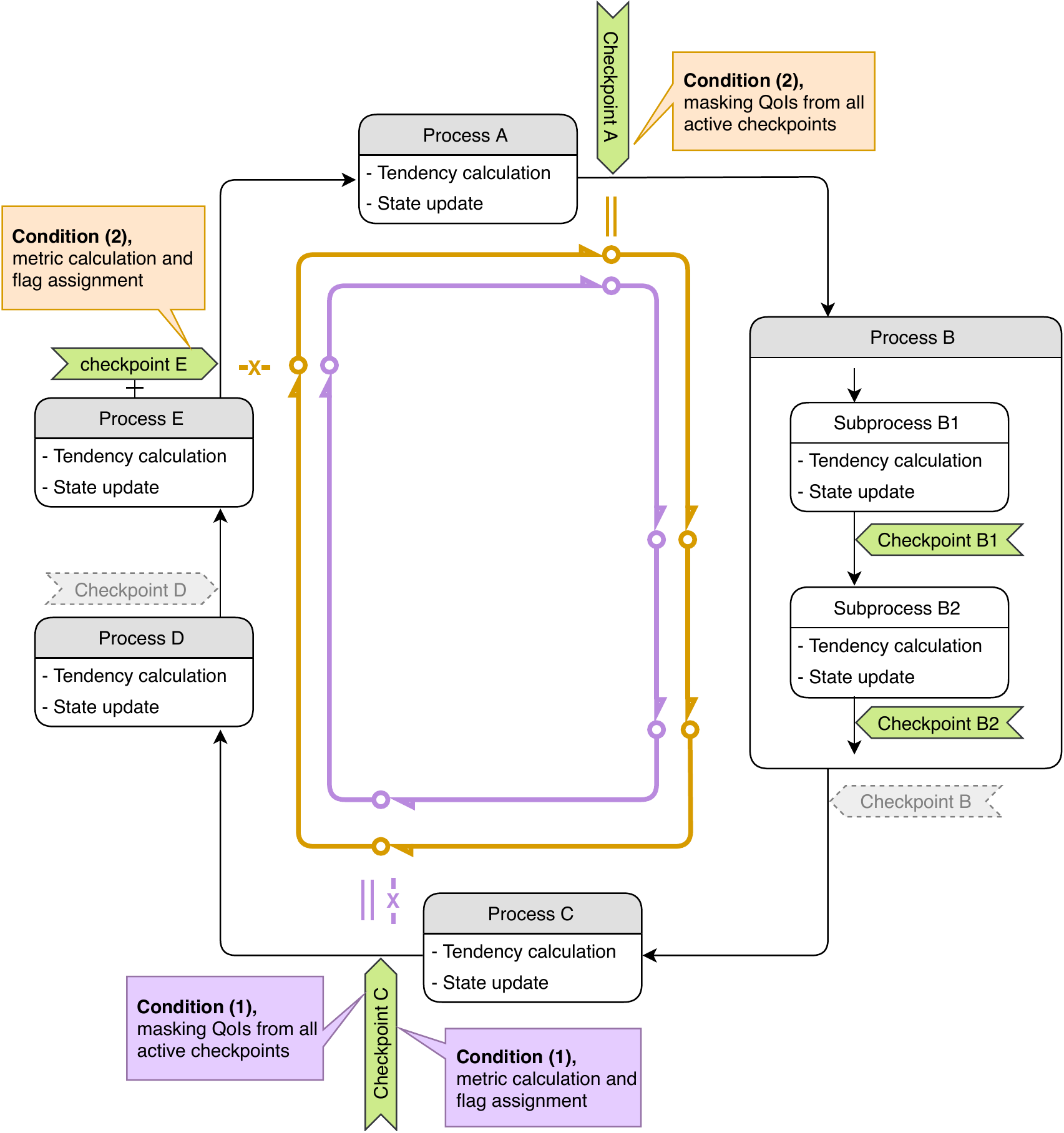}
\caption{
A schematic showing two sampling conditions
indicated in  brown and purple.
The ``-X-'' marks indicate locations in the time integration cycle 
where the condition metrics are evaluated. 
The double-bars indicate the end of validity of the evaluated sampling conditions.
More details can be found in Sect.~\ref{sec:start_and_end_of_time_step}.
Like in Fig.~\ref{fig:checkpoints}, 
green tags are active checkpoints being monitored by the tool.
Gray tags with dashed borderlines are inactive checkpoints,
which are ignored in the simulation.
}
\label{fig:conditions}\vspace{5mm}
\end{figure*}%

\subsubsection{Time window of validity of an evaluated condition}
\label{sec:start_and_end_of_time_step}

Our tool is designed 
to evaluate a sampling condition once per each 
$\Delta t_{\rm CPLmain}$ at a user-specified checkpoint X
and can monitor QoIs at multiple checkpoints within 
$\Delta t_{\rm CPLmain}$.
By default, the masking resulting from 
a condition evaluated at checkpoint X
is applied retrospectively to all active checkpoints from X 
until just before the previous encounter of X (i.e., X in the previous time step).
This is illustrated by condition (1) shown in purple 
in Fig.~\ref{fig:conditions}, where
the sampling condition is evaluated at checkpoint C
and the masking is applied retrospectively to 
checkpoints B2, B1, A, and E.

To provide more flexibility, our tool also allows 
the user to specify a different checkpoint 
as the end-of-validity mark for a sampling condition,
which we indicate with double-bars in Fig.~\ref{fig:conditions}. 
A hypothetical example is given as condition (2) shown in brown
in the figure. 
There, the end-of-validity mark (brown double-bar)
is placed at checkpoint A while the sampling 
condition is evaluated at checkpoint E. 
The masking determined at E is applied to 
E and the subsequent checkpoint A,
as well as retrospectively to checkpoints C, B2, and B1
before E.
An example from EAMv1 showing such 
a usage can be found in Sect.~\ref{sec:use_case_RHI}.

\subsection{Multiple sampling conditions in one simulation}
\label{sec:multiple_conditions}

A single sampling condition is defined by a combination of
(i) a metric, (ii) a threshold, (iii) 
a comparison type, 
(iv) a tolerance if the comparison type is ``equal to'', 
(v) a condition-evaluation checkpoint, and 
(vi) an end-of-condition-validity checkpoint.
Changing any of the elements will result in 
a new sampling condition.
Our tool allows for multiple conditions to be used 
in a single simulation (cf.~Fig.~\ref{fig:conditions}).

For software simplicity, 
the current implementation only allows one and the same 
set of QoIs and checkpoints to be monitored under all sampling conditions.
In the example illustrated in Fig.~\ref{fig:conditions} where 
two conditions ((1) and (2))
and five active checkpoints (A, B1, B2, C, and E) are activated, 
let us assume the user has chosen to monitor four QoIs, 
$T$, $q_v$, $u$, and $v$.
The same four QoIs and five checkpoints will be 
monitored for both sampling conditions.
The current implementation does not allow, for example,
monitoring only $T$ and $q_v$ at checkpoint A and C under condition (1)
and only $u$ and $v$ at checkpoints A, B1, and B2 under condition (2),
although this kind of flexibility can be considered for future versions of CondiDiag if needed.

\subsection{Mass-weighted vertical integral of QoIs}
\label{sec:x_dp}

For spatially 3D QoIs defined at layer midpoints or 
as cell averages, the vertical integral weighted by air mass 
can be calculated during the time integration 
and then conditionally sampled and written out as 2D variables.
This applies to both field values and their increments.

One note of caveat is that in EAM's physics parameterizations, 
the mixing ratios of water species
(vapor, cloud liquid and ice, rain and snow) are defined  
relative to the mass of moist air (i.e., dry air plus water vapor)
while the mixing ratios of aerosols and chemical gases 
are defined with respect to dry air. Our tool expects the 
user to specify which kind of air mass (moist or dry)
should be used for each QoI when vertical integrals is requested. 

Furthermore, while the physics parameterizations suite 
uses different mixing ratio definitions for different tracers, 
the numerical schemes for large-scale transport assume 
all tracers concentrations
are ``wet'' mixing ratios (i.e., defined relative to moist air),
and the wet-to-dry and dry-to-wet conversions occur during 
each model time step of size $\Delta t_{\rm CPLmain}$.
Therefore, our tool also allows a user to 
(1) specify the type of air mass to be used for vertical integral 
at each checkpoint, and 
(2) clarify  
whether the air mass type specified 
for the QoI or the checkpoint should take precedence
for each QoI-checkpoint combination.
The corresponding namelist parameters are explained in Sect.~\ref{sec:namelist_integral}.

\section{Implementation in EAMv1}
\label{sec:implementation}

This section explains how the design features described 
in Sect.~\ref{sec:design} are implemented in EAMv1. 
We start with an overview of the new Fortran modules   
added specifically for the tool (Sect.~\ref{sec:CondiDiag_modules}),
a general-purpose diagnostics module (Sect.~\ref{sec:diagnostics_module}),
and a summary of the changes 
made to the original EAMv1 code 
(Sect.~\ref{sec:code_changes_in_original_EAM}).
We keep these sections brief but provide two versions 
of the EAMv1 code on Zenodo corresponding to the GitHub commits before and after 
the implementation of CondiDiag1.0, 
so that readers can review the details of the code changes if needed.

\subsection{CondiDiag-specific new modules}
\label{sec:CondiDiag_modules}

Four new modules are added to define data structures and 
support key functionalities of our diagnostic tool.

\subsubsection{Data structure module}
\label{sec:data_structure_module}

The module \texttt{conditional\_diag} contains definitions of the basic data structures
used by our tool and subroutines for initializing the corresponding Fortran variables. 

A Fortran variable \texttt{cnd\_diag\_info} of the derived type
\texttt{cnd\_diag\_info\_t} contains the metadata that describes the 
user's conditional sampling strategy and/or budget analyses configuration.
A namelist \texttt{conditional\_diag\_nl} (cf.~Sect.~\ref{sec:namelist})
is also defined in this module, and
a subroutine \texttt{cnd\_diag\_readnl} parses the user's namelist
input and populates the information to 
{cnd\_diag\_info}.

A second derived type \texttt{cnd\_diag\_t} is defined for storing 
the values of the metrics, flags, as well as QoI
field values and increments. 
The corresponding Fortran variable is an array named \texttt{phys\_diag};
the array is defined in a different module (explained in Sect.~\ref{sec:phys_diag_array}).
The subroutines that allocate memory for elements of \texttt{phys\_diag}
and their components are included in module \texttt{conditional\_diag}.

\subsubsection{Key algorithm module}
\label{sec:key_algorithm_modules}

The module \texttt{conditional\_diag\_main} contains the key subroutine
of our tool named \texttt{cnd\_diag\_checkpoint}, which obtains the values of 
the condition metrics and QoIs, calculates the QoI increments, 
evaluates the sampling conditions, applies conditional sampling, 
and transfers the sampled fields to the output-handling infrastructure of EAM.
Examples showing how the subroutine is invoked in EAM is explained   
in Sect.~\ref{sec:checkpoint_call}.

As mentioned earlier in Sect.~\ref{sec:checkpoint_concept}, the condition metrics and QoIs
can be existing components of EAM's state variable, physics buffer, and
the \texttt{cam\_in} and \texttt{cam\_out} data structures. 
Taking air temperature as an example, the values are retrieved in 
subroutine \texttt{get\_values} in module
\texttt{conditional\_diag\_main} by

\begin{scriptsize}
\begin{verbatim}
case('T')
  arrayout(1:ncol,:) = state%t(1:ncol,:)
\end{verbatim}
\end{scriptsize}
Condition metrics and QoIs can also be 
physical quantities that need to be calculated from components of EAM's existing 
data structures. For example, the relative humidity with respect to ice 
is obtained by

\begin{scriptsize}
\begin{verbatim}
case ('RHI')
  call relhum_ice_percent(  &
       ncol, pver,          &! intent(in)
       state%t(:ncol,:),    &! intent(in)
       state%pmid(:ncol,:), &! intent(in)
       state%q(:ncol,:,1),  &! intent(in)
       arrayout(:ncol,:)    )! intent(out)
\end{verbatim}
\end{scriptsize}

In these examples, ``T'' and ``RHI'' need to be unique names 
within the module \texttt{conditional\_diag\_main};
these will also be the metric or QoI names that the users refer to 
in the namelist \texttt{conditional\_diag\_nl} (cf.~Sect.~\ref{sec:namelist}).
The currently implemented metric and QoI names
are listed in Table~\ref{tab:available_metrics_and_QoIs} in
Appendix~\ref{apndx:available_metrics_and_QoIs}.
Additional metrics and QoIs can be added 
following the existing examples. 
We note that some of the variable names in 
Table~\ref{tab:available_metrics_and_QoIs} coincide with
EAM's standard history variable names but the coincidence 
has no significance. 
Because a QoI can be monitored at different checkpoints 
and under different conditions, those different combinations 
will each correspond to a distinct variable name in the history files,
as explained in Sect.~\ref{sec:cnd_output}.

\subsubsection{History output module}
\label{sec:cnd_output}

The module \texttt{conditional\_diag\_output\_utils} is responsible for
adding the following items to EAM's master list of history output variables:
\begin{itemize}

\item the conditionally sampled metric field named with the pattern 
      \texttt{cnd<index>\_<metric\_name>} where \texttt{<index>} 
      is a two-digit number (e.g., \texttt{cnd01\_T} if the first 
      sampling condition uses air temperature as the metric);
      
\item the flag field (see Sect.~\ref{sec:metric_and_flag_field}) 
      named \texttt{cnd<index>\_<metric\_name>\_flag};

\item one output variable corresponding to each QoI at each active checkpoint
      under each sampling condition,  named with the pattern 
      \texttt{cnd<index>\_<QOI\_name>\_<checkpoint\_name>}.
      For example, \texttt{cnd01\_CLDLIQ\_DYNEND} is
      the stratiform cloud liquid mixing ratio monitored at 
      checkpoint DYNEND under condition 1. 
      If increments of the QoI are calculated and archived, 
      these will be named similar to the QoIs but with a suffix \texttt{\_inc} 
      append, 
      e.g., \\ \texttt{cnd01\_CLDLIQ\_DYNEND\_inc} for the increment of CLDLIQ 
      at checkpoint DYNEND under condition 1.

\item If the mass-weighted vertical integral is requested for a QoI, then
      a suffix \texttt{\_v} will be appended to the QoI name. For example,
      \texttt{cnd01\_CLDLIQ\_v\_DYNEND} is the 
      column burden of CLDLIQ at checkpoint DYNEND under condition 1 
      and \texttt{cnd01\_CLDLIQ\_v\_DYNEND\_inc} is the corresponding increment.

\end{itemize}
We expect that users of our tool should not need to touch 
the \texttt{conditional\_diag\_output\_utils} module
unless they want to 
revise the naming conventions for variables in the history files.

It is worth noting that
for any of the output variables added by our tool, EAM's standard 
history output functionalities apply (cf.~Sect.~\ref{sec:history_output_from_EAMv1}). 
For example, 
each variable can be added to or excluded from one or multiple history tapes
and be written out at the user-specified frequencies.
For temporal statistics, both instantaneous and time-averages
can be used in the current implementation. Maximum and minimum values
etc. need to be used with care as unselected grid cells are
filled with zeros. In future versions, we will consider 
allowing the user to specify what missing values should be assigned
to each QoI.

\subsubsection{Restart module}
\label{sec:restart_module}

Because our diagnostic tool uses its own data structure, 
new subroutines have been included to add additional contents to  
EAM's restart files. These subroutines
are placed in the module \texttt{conditional\_diag\_restart}.
As long as users do not change the data structures 
defined in module \texttt{conditional\_diag}, there should 
be no need to touch the restart module even if they
add new metrics and QoIs to the key algorithm modules
\texttt{conditional\_diag\_main} and \texttt{misc\_diagostics}.

\subsection{General-purpose diagnostics module}
\label{sec:diagnostics_module}

In case a user provides their own subroutines 
to calculate new metrics or QoIs, like \texttt{relhum\_ice\_percent} 
in the code snippet above,
we recommend those subroutines be placed in the module 
\texttt{misc\_diagnostics} rather than in \\
\texttt{conditional\_diag\_main},
because we view those user-provided subroutines
as general-purpose diagnostic utilities that 
could also be used by other parts of EAM (e.g., some parameterizations).

\subsection{Other code changes in EAMv1}
\label{sec:code_changes_in_original_EAM}

Other than adding the five modules explained in 
Secitons~\ref{sec:CondiDiag_modules}
and \ref{sec:diagnostics_module}, the implementation of our tool in
EAMv1 only involved a very small number of code changes,
as described below.

\subsubsection{The \texttt{phys\_diag} array and its elements}
\label{sec:phys_diag_array}

Our tool has the derived data type 
\texttt{cnd\_diag\_t} for storing
values of the condition metrics, flags, and QoI
field values as well as increments and vertical integrals, 
while the data storage closely follows 
the handling of EAM's model state variable:
a Fortran array of type \texttt{cnd\_diag\_t} is declared as an array of rank one
with the different array elements corresponding to different grid-cell chunks
handled by the same CPU.
The subroutines \texttt{tphysbc} and \texttt{tphysac}
that organize the invocation 
of individual parameterizations and their coupling 
operates on a single grid-cell chunk;
the scalar variable of type \texttt{cnd\_diag\_t}
in \texttt{tphysbc} and \texttt{tphysac} is named \texttt{diag}.


\subsubsection{Checkpoints}
\label{sec:checkpoint_call}

Most of the checkpoints listed in Tables~\ref{tab:checkpoints_bc} and \ref{tab:checkpoints_ac}
are added to subroutines \texttt{tphysbc} and \texttt{tphysac}
by inserting code lines like

\begin{scriptsize}
\begin{verbatim}
  call cnd_diag_checkpoint( diag, &! intent(inout)
       'DYNEND', state, pbuf,     &! intent(in)
       cam_in, cam_out            )! intent(in)
\end{verbatim}
\end{scriptsize}
where \texttt{diag} is the variable of type \texttt{cnd\_diag\_t} 
explained in Sect.~\ref{sec:phys_diag_array} and 
"DYNEND" is the unique string identifying this checkpoint,
\texttt{state}, \texttt{cam\_in} and  \texttt{cam\_out} are scalar 
variables of derived types declared in the original EAM code.

Checkpoints have also been included in the stratiform cloud 
macrophysics driver subroutine \texttt{clubb\_tend\_cam} in the form of, e.g.,

\begin{scriptsize}
\begin{verbatim}
  call cnd_diag_checkpoint( diag,     &! intent(inout)
       'CLUBB'//char_macmic_it,       &! intent(in)
       state1, pbuf, cam_in, cam_out  )! intent(in)   
\end{verbatim}
\end{scriptsize}
where the character string \texttt{char\_macmic\_it} labels 
the sub-steps within a full time step $\Delta t_{\rm CPLmain}$.
It is worth emphasizing that 
\texttt{state1} (instead of \texttt{state})
is referred to in the code snippet quoted above
because \texttt{state1} is the atmospheric state variable
that is sequentially updated by various subprocesses in
\texttt{clubb\_tend\_cam}.

\section{User's guide}
\label{sec:users_guide}

The new tool is expected to be useful for a wide range of simulations 
routinely performed by the model developers and users, 
including debugging simulations that are a few time steps long, 
short few-day simulations for preliminary testing 
or weather forecast style simulations for comprehensive evaluations of 
the model physics following protocols like Transpose-AMIP  
\citep[e.g.,][]{phillips:2004,williams:2013_transpose_amip_II,williamson:2005,martin:2010,Xie:2012,ma:2013a,ma:2014}, 
as well as more traditional multi-year to multi-decade simulations. 

To obtain process-level understanding of model behavior, 
it can be useful to use the new tool in an iterative manner.
For example, for a study like \citet{Zhang_et_al:2018} 
where one needs to identify model processes that result in 
negative values of specific humidity, 
we can start by carrying out a few-day or one-month simulation 
with unconditional sampling, choosing a large number of checkpoints
to monitor model processes that are expected to affect humidity 
or might inadvertently do so because of computational artifacts 
or code bugs. We let the tool diagnose and archive time averages
of the specific humidity and its increment at these checkpoints
to get a sense of typical values of the state variable 
and identify sources and sinks of moisture.
In a second step of investigation, we eliminate 
from the previous selection any checkpoints that have been confirmed
to not see humidity change in any grid cell or time step in 
the few-day or one-month simulation.
From the shorter list, we can pick one or multiple 
model processes as suspected culprits of negative 
specific humidity. If $m$ suspects are selected for further 
investigation, then $m$ sampling conditions can be specified 
in the next simulation, 
all using $q_v < 0$ as the sampling criterion
but each evaluated after a different suspect.
We also select some QoIs
(e.g., temperature, specific and relative humidity,
wind, total cloud fraction, cloud liquid and ice 
mixing ratios, etc.), to be monitored both 
right before and right after the model processes that
are suspected to cause negative water vapor.
We can request both the field values and increments
of these QoIs to be archived, as time averages or 
instantaneous values (or both). This second step 
might provide useful clues of the typical meteorological 
conditions under which negative water vapor 
is predicted in the model. 
If pathological conditions are identified, 
then we can carry additional simulations using 
relevant sampling conditions to further investigate 
the causes of those pathologies.

This section explains how investigations described above
can be performed using our tool. 
We first present a typical workflow in Sect.~\ref{sec:user_workflow}
to illustrate the steps that a user needs to go through  
when designing an analysis and setting up an EAM simulation using our tool.
We then explain the namelist parameters of our tool in Sect.~\ref{sec:namelist}.

\begin{figure*}[t]
\includegraphics[height=16cm]{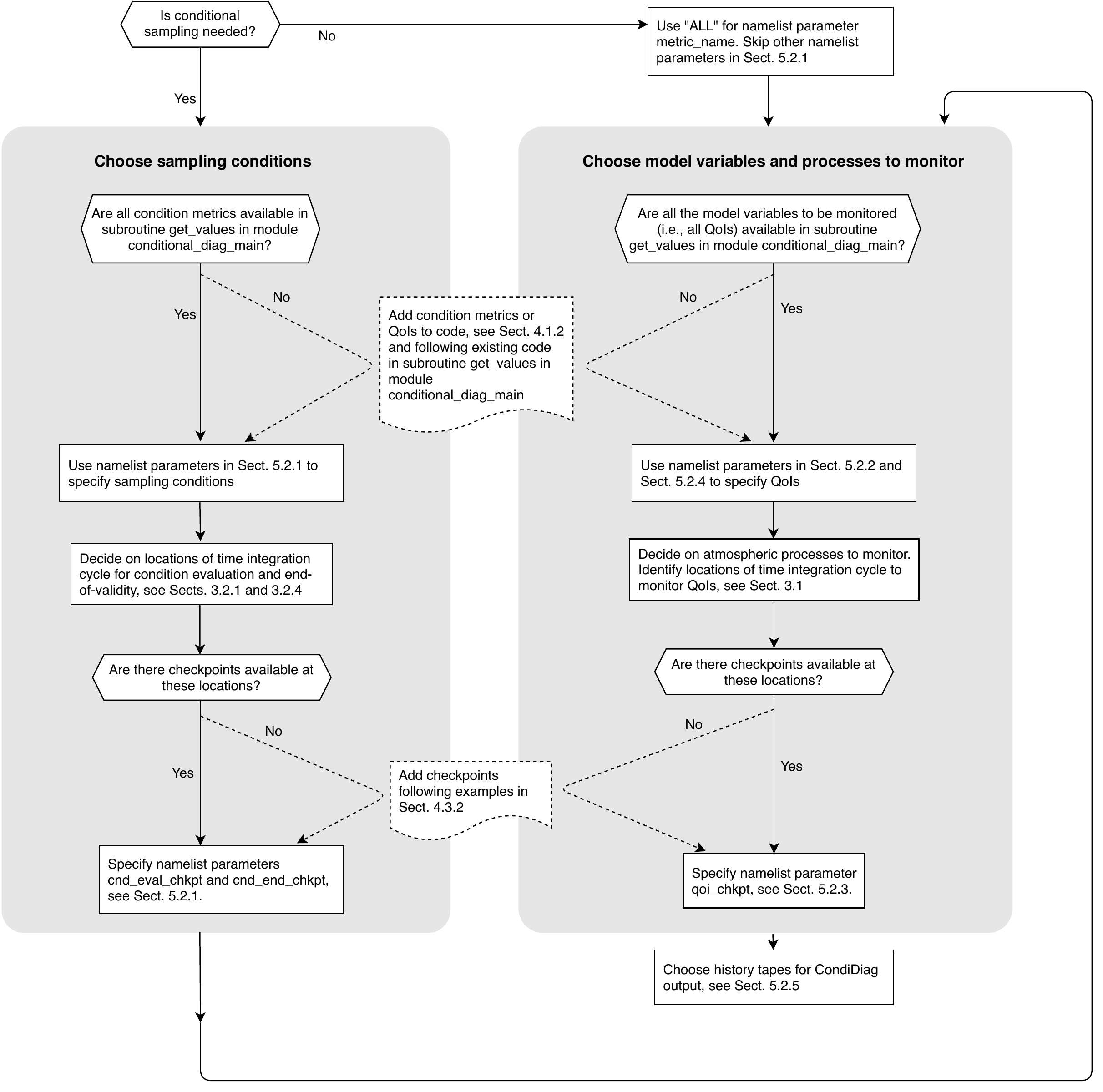}
\caption{
A schematic showing a typical workflow 
to illustrate the steps that a user needs to go through  
when setting up an EAM simulation using our tool.
Dashed lines indicate places where code changes or additions 
are needed from the user.
}
\label{fig:user_workflow}
\end{figure*}%

\subsection{User workflow}
\label{sec:user_workflow}

The schematic in Fig.~\ref{fig:user_workflow} summarizes 
the steps to take when designing a composite or budget analysis using our tool.
It also points to relevant concepts explained in 
earlier sections and namelist parameters explained below. 

\subsection{Namelist \texttt{conditional\_diag\_nl}}
\label{sec:namelist}

Users specify their 
conditional sampling  and budget analysis strategy
via the namelist 
\texttt{conditional\_diag\_nl}, which
consists of five groups of parameters.

\subsubsection{Specifying sampling conditions}
\label{sec:namelist_conditions}

For specifying sampling conditions, we have
\begin{itemize}

\item \texttt{metric\_name}, a character string array containing
the names of the condition metrics to be used in a simulation;

\item \texttt{metric\_nver}, an integer array specifying 
the number of vertical levels of each metric is defined with. 
This is meant to help distinguish physical quantities that
(1) have no vertical dimension, (2) are defined at 
layer mid-points, and (3) are defined at layer interfaces. 
Valid values for {metric\_nver}
are 1, \texttt{pver} (e.g., 72), and \texttt{pverp} (e.g., 73),
where \texttt{pver} and \texttt{pverp} are EAM's variable
names for the number of vertical layers and interfaces, respectively.
 
\item \texttt{metric\_cmpr\_type}, an integer array specifying
the types of comparison to be used for each condition  (one entry per condition):
0 for ``equal to within a tolerance'', 1 for ``greater than'', 
2 for ``greater than or equal to'', -1 for ``less than'',
and -2 for ``less than or equal to'';

\item \texttt{metric\_thereshold}, a double-precision floating-point 
array specifying the threshold values that the metrics will be 
compared to (one threshold for each condition);

\item \texttt{metric\_tolerance}, a double-precision floating-point 
array specifying the tolerances for conditions with comparison type 0
(one tolerance for each condition; the value will have no effect for 
conditions with a non-zero comparison type);

\item \texttt{cnd\_eval\_chkpt}, a character string array 
specifying at which checkpoints the conditions will be evaluated
(see Sect.~\ref{sec:defining_a_condition}; 
one checkpoint for each condition).

\item \texttt{cnd\_end\_chkpt}, a character string array 
specifying the checkpoints defining the end of validity of an evaluated condition
(see Sect.~\ref{sec:start_and_end_of_time_step}; one checkpoint per condition). 
If not specified by user, the end-of-time-step checkpoint
will be set to the condition-evaluation checkpoint (\texttt{cnd\_eval\_chkpt}).

\end{itemize}

\subsubsection{Specifying monitored model variables}
\label{sec:namelist_QoIs}

The QoIs to be monitored are specified via a character string 
array \texttt{qoi\_name}, and the number of vertical levels of 
each QoI is given by the integer array \texttt{qoi\_nver}.
If no QoIs are specified but some sampling condition have been
chosen, then conditional sampling will only be applied to the metrics.

The monitoring of QoI field values are turned on by  
the logical scalar \texttt{l\_output\_state}. 
A second logical scalar, \texttt{l\_output\_incrm}, 
is used to turn on or off the monitoring of QoI increments. 
Users' choice for the two switches will be applied to all QoIs.

\subsubsection{Choosing checkpoints}
\label{sec:namelist_checkpoints}

The checkpoints at which the QoIs will be monitored are specified by 
a character string array \texttt{qoi\_chkpt}.
The sequence in which they are mentioned in the namelist has no significance.
Note that the same checkpoints will be applied to all QoIs.
Also note that if the user specifies a checkpoint name that does not 
match any checkpoint implemented in the code (e.g., because of a typographical error),
then our tool will act as if the wrong checkpoint is an inactive one
- in the sense that it will get ignored when the tool 
attempts to obtain QoI field values and calculate increments
as the simulation proceeds;
the history files will contain output variables corresponding to 
the incorrect checkpoint name but those output variables 
will contain zeros.


\subsubsection{Turning on vertical integral}
\label{sec:namelist_integral}

The calculation of mass-weighted vertical integrals of QoIs are enabled 
by two integer arrays \texttt{qoi\_x\_dp} and \texttt{chkpt\_x\_dp}:

\begin{itemize}
\item 
\texttt{chkpt\_x\_dp} is expected to be specified in relation to 
\texttt{qoi\_chkpt} 
(one value of \texttt{chkpt\_x\_dp} for each checkpoint. 
A value of 1 tells our tool the mass of moist air
should be used while a value of 2 indicates dry air mass should be used. 
Any other values assigned to \texttt{chkpt\_x\_dp} 
will be interpreted as no specification.

\item \texttt{qoi\_x\_dp} is expected to be specified in relation to 
\texttt{qoi\_name}, i.e., one value of \texttt{qoi\_x\_dp} for each QoI.
0 is interpreted as no integral; the QoI will be sampled and written out
as a 3D field. If 1 (moist) or 2 (dry) are selected, 
the corresponding (moist or dry) air mass will be used for vertical integral of
that QoI at {\it all} active checkpoints in the simulation.
If a value of 101 (moist) or 102 (dry) is used, 
the the corresponding air mass will be used for that QoI 
at all active checkpoints {\it except} where
\texttt{chkpt\_x\_dp} indicates a different specification for a checkpoint.
For example, let us assume 
we set \texttt{qoi\_x\_dp}~=~102
for the coarse mode dust mass mixing ratio;
we choose to monitor checkpoints A, B, and C
and set \texttt{chkpt\_x\_dp}~=~0,0,1.
Then, when our tool calculates the coarse mode dust burden, 
the dry air mass will be used for checkpoints A and B 
while the moist air mass will be used for checkpoint C.
In other words, a value of \texttt{qoi\_x\_dp} larger than 100
means using $\mod(qoi\_x\_dp,100)$ in general but giving \texttt{chkpt\_x\_dp} 
precedence when the latter is set to non-zero at a checkpoint.

\end{itemize}
If the user wishes to monitor both a 3D QoI and its vertical integral,
they can specify the same QoI twice in \texttt{qoi\_name}, 
and then set one of the corresponding \texttt{qoi\_x\_dp} element to 0 and 
the other to an appropriate value to request vertical integral.
An use case example is provided in Sect.~\ref{sec:use_case_dust_budget}.

\subsubsection{Turning on history output}
\label{sec:namelist_outputl}

A user might want to write out multiple 
copies of the conditional diagnostics or budget diagnostics 
to different history files
that correspond to different output frequencies and/or temporal averaging.
To facilitate such needs, the integer array 
\texttt{hist\_tape\_with\_all\_output} specifies 
which history files will contain 
the full set of output variables from our tool. 
For example, \texttt{hist\_tape\_with\_all\_output} = 1, 3
will include the output to the h0 and h2 files.
Again, we note that the standard output 
functionalities in EAM explained in Sect.~\ref{sec:history_output_from_EAMv1} still apply.

\subsection{Using unconditional sampling}

One of the main motivations for creating our tool is 
to facilitate budget analysis. 
If an analysis is to be carried out for the entire computational domain 
and all time steps, then a special metric named \texttt{ALL} can be used.
In such a case, the user can ignore (skip) 
the other namelist parameters in \ref{sec:namelist_conditions}.
When \texttt{ALL} is used, 
the condition evaluation will be skipped
during the model's integration (see example in Sect.~\ref{sec:use_case_dust_budget}).
Another way to use unconditional sampling is to 
specify a condition that will always be fulfilled 
(e.g., relative humidity higher than -1\%)
a use case example is provided in Sect.~\ref{sec:use_case_RHI}).

\section{Use case examples}
\label{sec:examples}

This section demonstrates the usage of the new tool 
using three concrete examples:

The first example is a global budget analysis without conditional sampling. 
It demonstrates how to request unconditional sampling 
and how to request that increments of model variables 
be calculated and archived as time averages.
This first example also demonstrates that with our tool, 
it is convenient to obtain both vertical profiles and 
vertical integrals of the budget terms.

The second example is a composite analysis without budget terms.
It demonstrates how to use multiple sampling conditions in the same simulation
and also shows that the tool can be used to perform 
a univariate probability distribution analysis. 

In the third example, the increment diagnosis and conditional sampling capabilities 
are combined to perform a conditional budget analysis. 
The example demonstrates how metrics and monitored QoIs can be chosen 
to be physical quantities that need to be calculated from the host 
model's state variables using user-provided subroutines.

The examples shown here use 1-month simulations of October 2009 
with monthly (or monthly and daily) output.
All simulations were carried out with active atmosphere and land surface 
as well as prescribed sea surface temperature and sea ice concentration,
at 1 degree horizontal resolution with out-of-the-box parameters 
and time integration configurations of EAMv1.

\subsection{A global budget analysis of dust aerosol mass mixing ratio and burden}
\label{sec:use_case_dust_budget}

The first example is a global 
dust aerosol mass budget analysis without conditional sampling.
The simulation is designed to provide insight into the atmospheric processes 
that change the burden (vertical integrals) of dust aerosols
in two size ranges (accumulation mode and coarse mode).
In particular, we are interested in dust emission, 
dry removal (i.e., sedimentation and dry deposition at the Earth's surface),
resolved-scale transport, 
subgrid-scale turbulent transport and 
activation (i.e., nucleation scavenging),
as well as the wet removal caused by precipitation collecting particles by impaction,
and resuspension caused by evaporation of precipitation.

\begin{table}[htbp]
\centering
\caption{
Namelist setup used in the dust budget analysis example  
in Sect.~\ref{sec:use_case_dust_budget}.
\label{tab:namelist_dust_budget}}
\scriptsize
\begin{tabular}{l}\tophline
\begin{minipage}{7.5cm}
\begin{verbatim}

metric_name = 'ALL' 

qoi_chkpt = 'CFLXAPP', 'AERDRYRM',
            'PBCINI',  'STCLD',    'AERWETRM'

qoi_name = 'dst_a1', 'dst_a1', 'dst_a3', 'dst_a3'
qoi_nver =  72,       72,       72,       72
qoi_x_dp =  0,        2,        0,        2

l_output_state = .false.
l_output_incrm = .true.

hist_tape_with_all_output = 1,  2
nhtfrq                    = 0, -24
mfilt                     = 1,  31

\end{verbatim}
\end{minipage}
\\
\bottomhline
\end{tabular}\vspace{5mm}
\end{table}

\begin{table}[htbp]
\centering\caption{
For the dust budget analysis example  
in Sect.~\ref{sec:use_case_dust_budget}:
atmospheric processes corresponding to 
increments diagnosed at the checkpoints
selected in the namelist shown in 
Table~\ref{tab:namelist_dust_budget}.
\label{tab:checkpoints_and_processes_dust_budget}
}
\footnotesize
\begin{tabular}{ll}\tophline
\textbf{Checkpoint} & 
\textbf{Atmospheric processes} \\\middlehline
 CFLXAPP   & Surface fluxes of aerosol and chemical tracers \\
 AERDRYRM  & Dry removal of aerosols\\
 PBCINI    & Resolved transport \\
 STCLD     & Turbulent mixing and aerosol activation \\
 AERWETRM  & Wet removal and resuspension of aerosols \\
\bottomhline
\end{tabular}\vspace{5mm}
\end{table}

\subsubsection{Simulation setup}

The namelist setup for this study is shown in Table~\ref{tab:namelist_dust_budget}.
Only one condition is specified: the special metric \texttt{ALL} 
is used to select the entire model domain and all time steps.

The QoI names \texttt{dst\_a1} and \texttt{dst\_a3} are EAM's tracer names 
for dust mass mixing ratio in the accumulation mode and coarse mode, respectively. 
Each tracer name is mentioned twice under \texttt{qoi\_name}, 
with corresponding \texttt{qoi\_x\_dp} values of
0 and 2, meaning that both the vertical distribution 
of the tracer and its column burden are monitored.
With \texttt{l\_output\_state} set to \texttt{.false.}
and \texttt{l\_output\_incrm} set to \texttt{.true.},
the tool captures the dust mass mixing ratio increments 
caused by the targetted atmospheric processes but not the mixing ratios.
Five checkpoints are chosen for monitoring the dust budget.
The corresponding atmospheric processes are listed in
Table~\ref{tab:checkpoints_and_processes_dust_budget}.
(We remind the users that, 
as shown in Fig.~\ref{fig:checkpoints},
the model processes that contribute to increments 
diagnosed at a checkpoint not only depends 
on where this checkpoint is located in the time integration cycle
but also where the previous active checkpoint is located.)

The full set of fields tracked by our tool are sent to 
output files 1 (the h0 file)  and 2 (the h1 file), 
with the h0 file containing monthly averages and
the h1 file containing daily averages.

\begin{figure}[htbp]
\includegraphics[height=19cm]{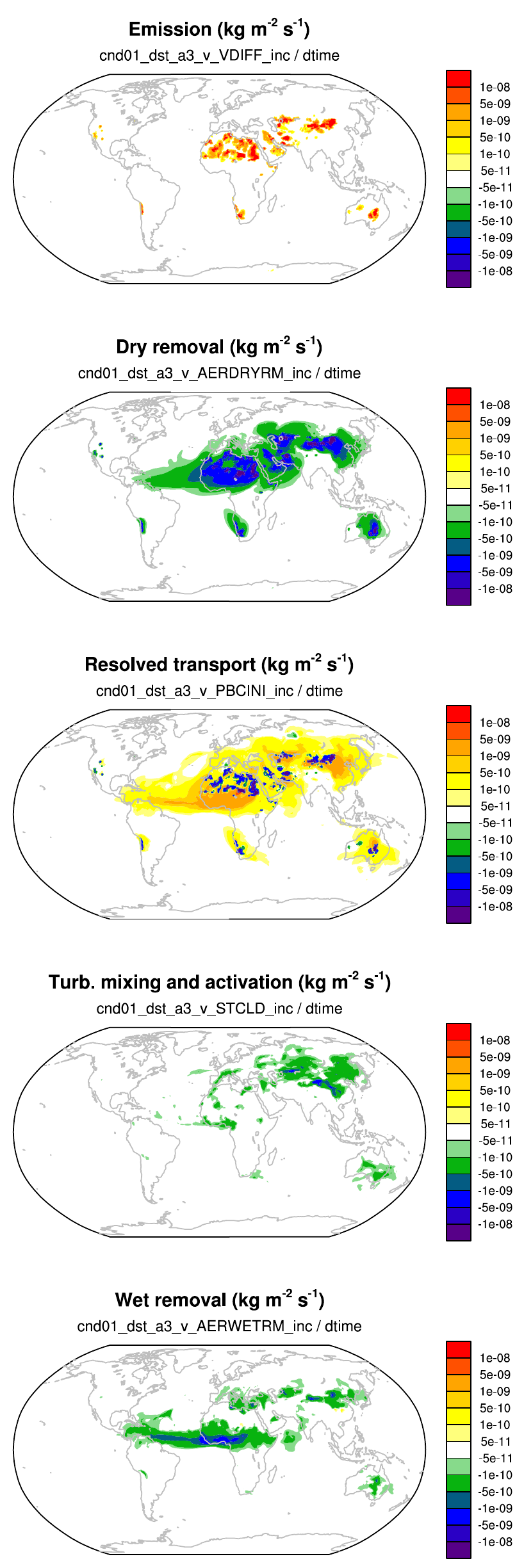}
\caption{\label{fig:dst_a3_monthly_mean_budget_maps}
One-month mean tendencies of the vertically integrated 
coarse mode dust burden (unit: kg~m$^{-2}$~s$^{-1}$)
attributed to different physical processes in EAMv1.
The expressions given in thin fonts below panel titles
indicate how the presented quantities are calculated from the
model's output variables.
}
\end{figure}

\begin{figure}[htbp]
\includegraphics[width=5.5cm]{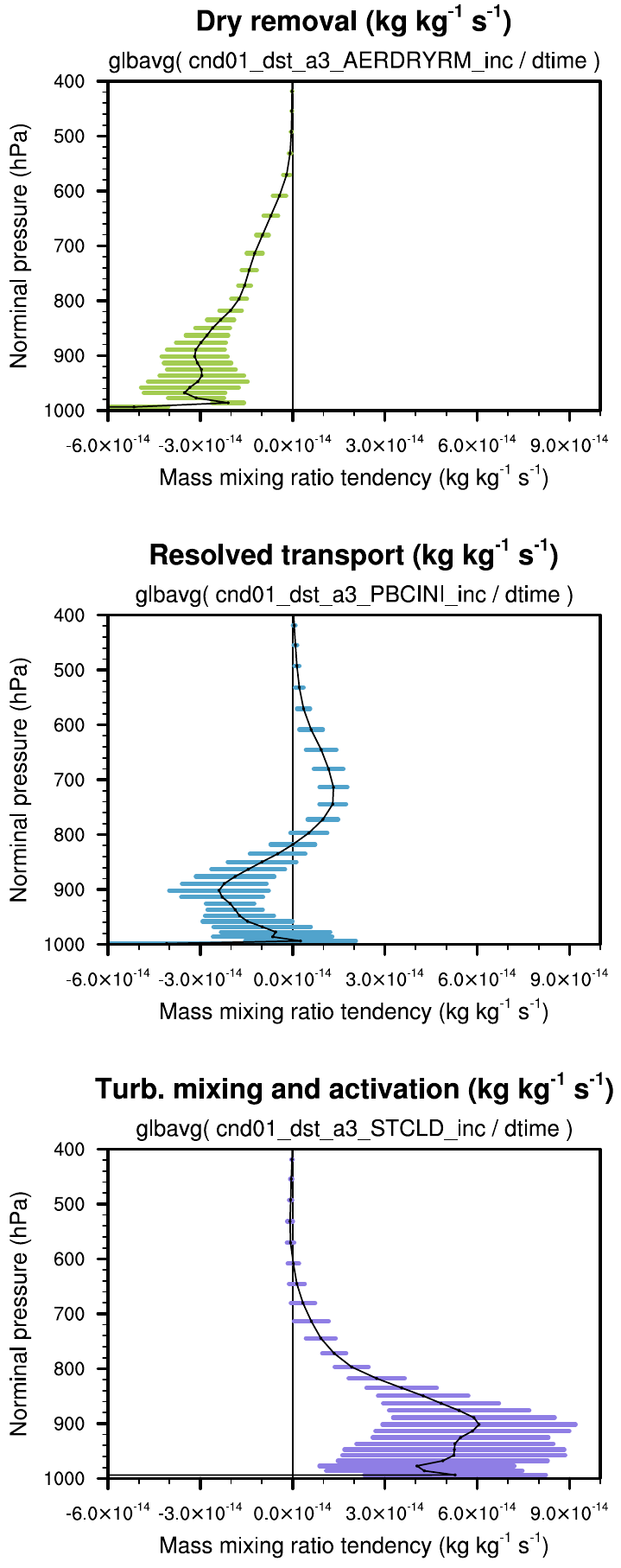}
\caption{\label{fig:dst_a3_dry_removal_profile}
Globally averaged vertical profile of  
the coarse mode dust mass mixing ratio tendencies   
(unit: kg~kg$^{-1}$~s$^{-1}$)
attributed to dry removal (top panel), resolved transport (middle panel),
as well as turbulent mixing and activation of aerosol particles (bottom panel). 
The black curves are monthly averages.
The lengths of the horizontal bars 
correspond to twice of the standard deviation of 
the daily averages.
The expressions given in thin fonts below panel titles
indicate how the global averages are calculated from the
model's output variables.
}
\end{figure}

\subsubsection{Results}

Figure~\ref{fig:dst_a3_monthly_mean_budget_maps}
shows a one-month mean geographical distribution of the 
sources and sinks of dust mass in the coarse mode (unit: kg~m$^{-2}$~s$^{-1}$).
The values shown are the output variables 
\texttt{cnd01\_dst\_a3\_v\_<checkpoint\_name>\_inc} in the h0 file
divided by $\Delta t_{\rm CPLmain}$~=~30~min.
Figure~\ref{fig:dst_a3_dry_removal_profile}
shows examples of the globally averaged vertical profiles of 
the coarse-mode dust mass mixing ratio tendencies 
(unit: kg~kg$^{-1}$~s$^{-1}$).
The black curves are monthly averages.
The colored horizontal bars indicate
variability of the daily averages derived 
from the 3D increment fields 
\texttt{cnd01\_dst\_a3\_<checkpoint\_name>\_inc} written to the h1 file.

\subsection{A composite analysis of sea salt emissions in relation to surface wind speed}
\label{sec:use_case_U10}

This example demonstrates the use of composite analysis 
(without budget terms) to provide insight into wind speed impacts 
on emission fluxes of sea salt aerosol in various size ranges. 
The intension is to examine the geographical distribution 
of sea salt emission fluxes under weak, medium, and strong
wind conditions and quantify their relative contributions
to the total emission fluxes.


\begin{table}[htbp]
\caption{\label{tab:namelist_U10}
Namelist setup used in 
the composite analysis presented 
in Sect.~\ref{sec:use_case_U10}.
}
\scriptsize
\begin{tabular}{l}\tophline
\\
\begin{minipage}{7cm}
\begin{verbatim}
metric_name      = 'U10', 'U10', 'U10', 'U10'
metric_nver      =  1,     1,     1,     1
metric_cmpr_type = -1,     0,     1,     1
metric_threshold =  5,     7.5,   10,   -1
metric_tolerance =  0,     2.5,   0,     0
cnd_eval_chkpt   = 'CHEM','CHEM','CHEM','CHEM'

qoi_chkpt = 'CHEM'

qoi_name = 'SFncl_a1', 'SFncl_a2', 'SFncl_a3'
qoi_nver =  1,          1,          1

l_output_state = .true.
l_output_incrm = .false.

hist_tape_with_all_output = 1
nhtfrq = 0
mfilt  = 1

\end{verbatim}
\end{minipage}
\\ \bottomhline
\end{tabular}
\end{table}

\begin{figure*}[t]
\includegraphics[width=15cm]{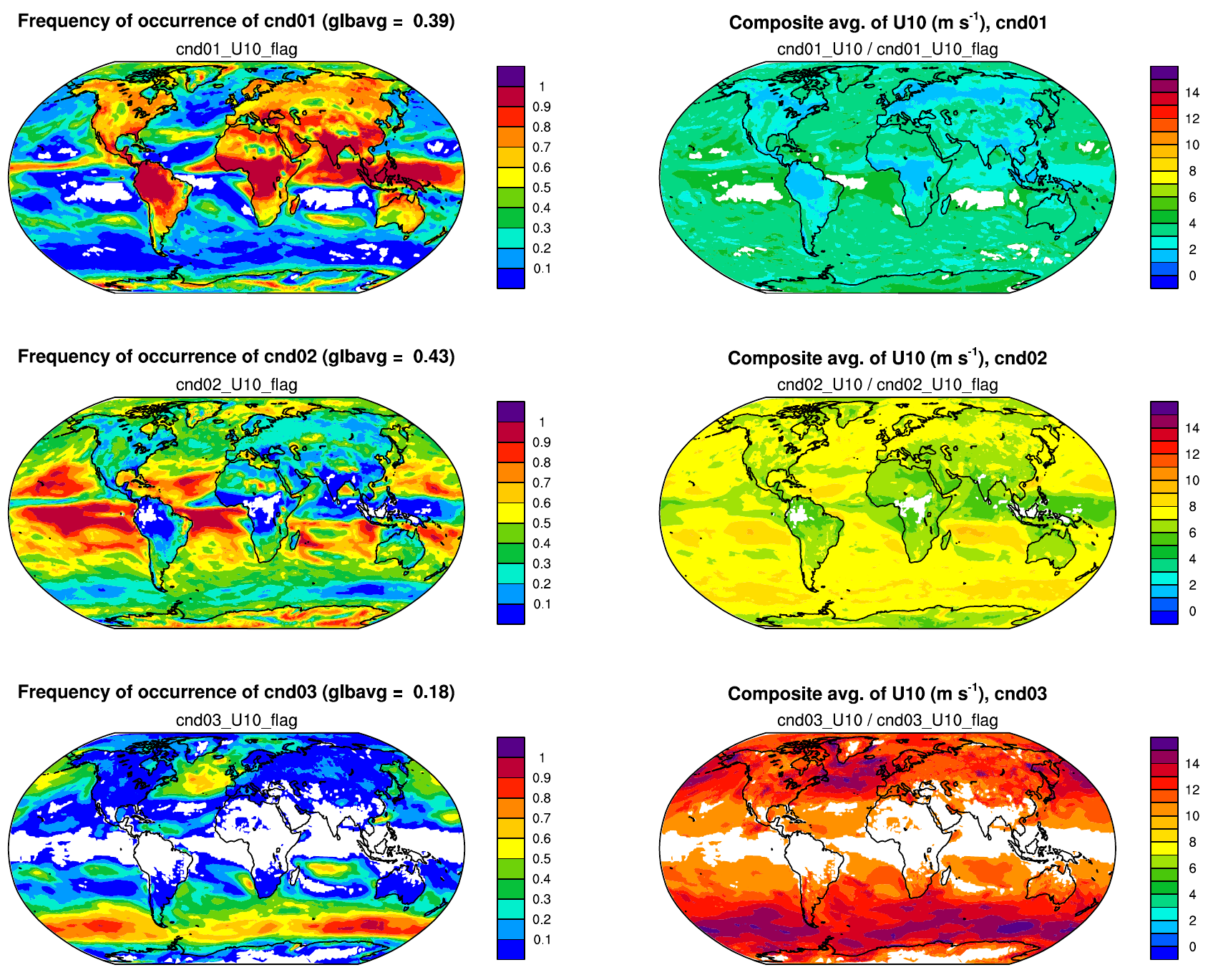}
\caption{\label{fig:U10_and_freq_maps}
Left column: geographical distributions of the frequency of occurrence 
of conditions 1--3 corresponding to 10~m wind speed (U10) 
$< 5$~m~s$^{-1}$ (top row),
between 5~~m~s$^{-1}$ and 10~m~s$^{-1}$ (middle row) and 
$> 10$~m~s$^{-1}$ (bottom row), respectively.
Right column: composite average of U10 under each condition.
White areas in the contour plots correspond to no occurrence of condition in 
the one-month simulation.
The expressions given in thin fonts below panel titles
indicate how the presented quantities are calculated from the
model's output variables.
}\vspace{5mm}
\end{figure*}

\begin{figure*}[t]
\includegraphics[width=15cm]{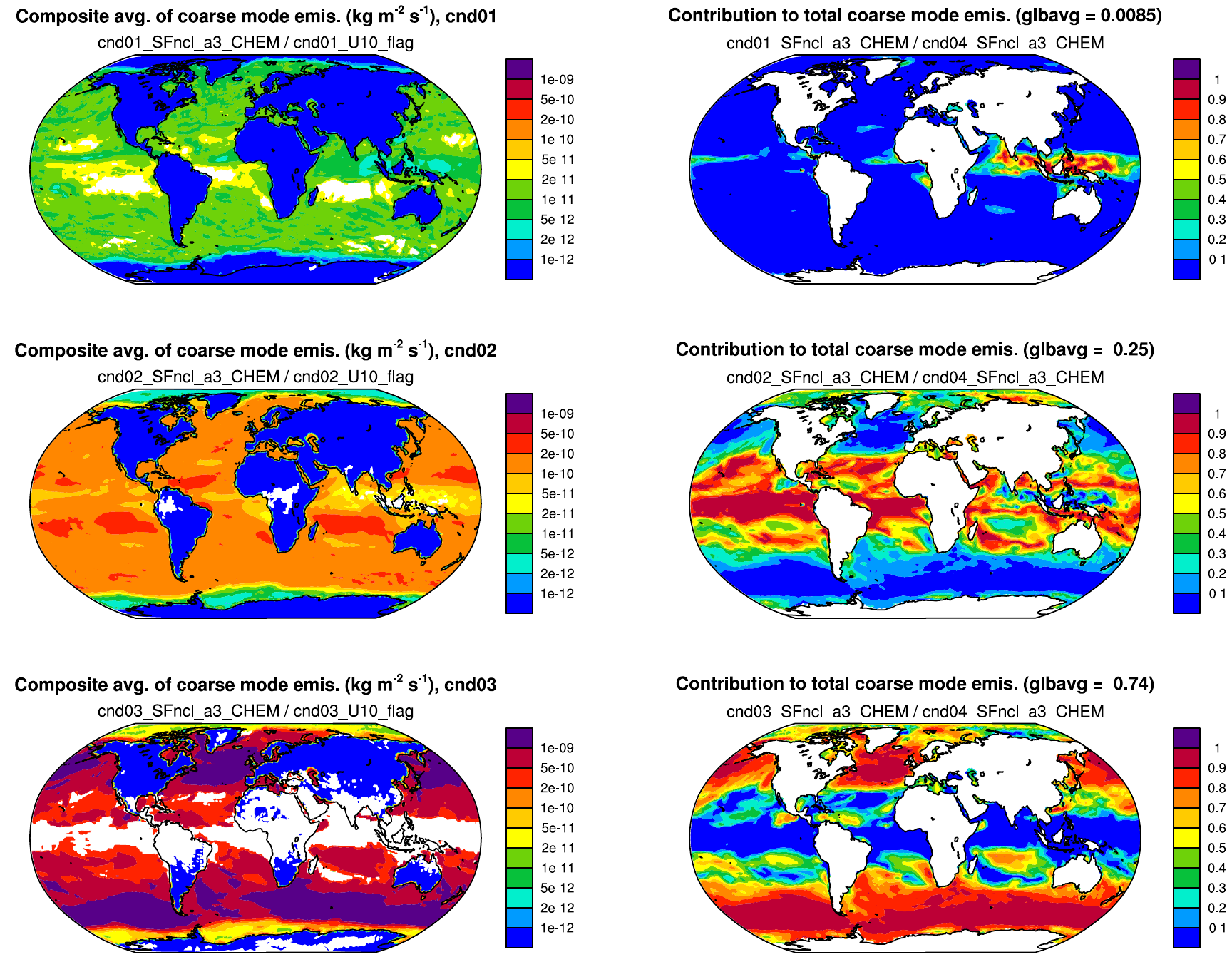}
\caption{\label{fig:sea_salt_emission_maps}
Left column: Composite average of coarse mode sea salt mass emission 
fluxes under conditions 1--3 corresponding to 10~m wind speed 
$< 5$~m~s$^{-1}$ (top row),
between 5~~m~s$^{-1}$ and 10~m~s$^{-1}$ (middle row) and 
$> 10$~m~s$^{-1}$ (bottom row), respectively.
Right column: contribute of each condition (1, 2, or 3) 
to the total coarse mode sea salt emission (condition 4).
White areas in the left panels are missing values caused 
by zero frequency of occurrence of the corresponding conditions.
White areas in the right panels are missing values caused 
by zero total coarse mode sea salt emission.
The expressions given in thin fonts below panel titles
indicate how the presented quantities are calculated from the
model's output variables.
}
\end{figure*}

\subsubsection{Simulation setup}

In EAMv1, the emission of sea salt aerosol
is parameterized with a scheme from \citet{Martensson_2003_JGR}
in which the emission flux is proportional to (U10)$^{3.41}$
with U10 being the wind speed (unit: m~s$^{-1}$) at 10~m above sea level
\citep{ZhangK_2016_GMD_subgrid_wind,Liu_2012_GMD_MAM}.

Four conditions are specified in the namelist setup shown in 
Table~\ref{tab:namelist_U10}. The first three divide 
the possible 10~m wind speed values into three ranges:
lower than 5~m~s$^{-1}$, 
between 5~m~s$^{-1}$ and 10~m~s$^{-1}$, and
higher than~10~m~s$^{-1}$.
The fourth condition uses the always-fulfilled criterion of 
U10~$>$~-1~m~s$^{-1}$
to select all grid points and time steps for comparison.

Three QoIs are monitored:  
SFncl\_a1, SFncl\_a2, SFncl\_a3, which are the surface 
mass fluxes of sea salt aerosol in the accumulation mode,
Aitken mode, and coarse mode, respectively. 
These variable names are EAMv1's standard tracer flux names.

U10 in EAMv1 is the grid-box average provided by 
the coupler (checkpoint \texttt{MCTCPL}).
The calculation of sea salt emissions is done in the atmosphere model
(checkpoint \texttt{CHEMEMIS}). 
U10 and the surface fluxes are calculated only once 
per time step $\Delta t_{\rm CPLmain}$ and their values remain
available as components of the derived-type Fortran variable 
called \texttt{cam\_in} (cf.~Table~\ref{tab:available_metrics_and_QoIs}).
Therefore, 
as long as we select any checkpoint at or after \texttt{MCTCPL} for assessing U10 
combined with any checkpoint at or after \texttt{CHEMEMIS},
and before \texttt{MCTCPL} for monitoring the surface fluxes,
the results will be equivalent.
In Table~\ref{tab:namelist_U10}, the same checkpoint \texttt{CHEM} is 
used for both namelist parameters 
\texttt{cnd\_eval\_chkpt} and 
\texttt{qoi\_chkpt}, as this is the checkpoint
right before the surface fluxes are used to update
aerosol tracer mixing ratios.

For output, variables from our tool are included in the h0 file as
monthly averages.

\subsubsection{Results}

Figure~\ref{fig:U10_and_freq_maps} presents geographical distributions 
of the frequency of occurrence of conditions 1--3 in the one-month
simulation (left column) and the corresponding composite averages of 
U10 (right column). While composite averages of U10 are shown for a sanity check,
the left panels indicate the different characteristic wind speed
associated with different surface types (land versus ocean) and cloud regimes
(e.g., deep convection active regions, trade cumulus regions,
and storm tracks).

Figure~\ref{fig:sea_salt_emission_maps} shows geographical distributions 
of the composite mean of the coarse mode sea salt mass emission fluxes 
under conditions 1--3 (left column) and the relative contribution of 
each condition to the total (all-condition) fluxes (right column).
Here, for demonstration purposes, we only chose 3 wind speed bins
and monitored sea salt mass fluxes.
If one refines the wind speed ranges (e.g., use 10 
to 20 bins), adds aerosol number fluxes to the QoIs, and 
adds the calculation of global averages to postprocessing, 
then diagrams like Fig.~5 in \citet{ZhangK_2012_HAM2} 
can be created to investigate the simulated 
relationship between wind speed and particle size distribution 
of the emissions but without having to write out a large amount 
of instantaneous model output.



\subsection{A conditional budget analysis for RHI}
\label{sec:use_case_RHI}

The third example demonstrates a combined use of the budget analysis and conditional sampling capabilities using our tool. The example also requires the calculation of a diagnosed quantity (the relative humidity with respect to ice, RHI) that is not a state variable, so additional routines are invoked to calculate it.  This quantity would vary before and after various processes (e.g., atmospheric dynamics, cloud microphysics, radiation etc) that operate on the atmospheric state, so it is sensitive to how and where it is calculated in the model, and it is also a function of the model sub-cycling.
%

\begin{table}[htbp]
\caption{\label{tab:namelist_RHI}
Namelist variables pertaining to the new diagnostic tool
used in the conditional RHI budget analysis presented 
in Sect.~\ref{sec:use_case_RHI}.
}
\scriptsize
\begin{tabular}{l}\tophline
\\
\begin{minipage}{7cm}
\begin{verbatim}
metric_name      = 'RHI',      'RHI'
metric_nver      =   72,         72
metric_cmpr_type =    1,          1
metric_threshold =  125,         -1
cnd_eval_chkpt   = 'CLDMAC01', 'CLDMAC01'
cnd_end_chkpt    = 'PBCDIAG',  'PBCDIAG'

qoi_chkpt = 'PBCDIAG',  'RAD',     'PACEND', 
            'DYNEND',   'DEEPCU',
            'CLDMAC01', 'CLDMIC01',
            'CLDMAC02', 'CLDMIC02',
            'CLDMAC03', 'CLDMIC03',
            'CLDMAC04', 'CLDMIC04',
            'CLDMAC05', 'CLDMIC05',
            'CLDMAC06', 'CLDMIC06'

qoi_name = 'RHI', 'Q', 'QSATI'
qoi_nver =  72,    72,  72

l_output_state = .true.
l_output_incrm = .true.

hist_tape_with_all_output = 1
nhtfrq = 0
mfilt  = 1

\end{verbatim}
\end{minipage}
\\\tophline
\end{tabular}
\end{table}

\begin{table}[htbp]
\centering\caption{
For the conditional RHI budget example presented  
in Sect.~\ref{sec:use_case_RHI}:
atmospheric processes corresponding to 
increments diagnosed at the checkpoints
selected in 
Table~\ref{tab:namelist_RHI}.
\label{tab:checkpoints_and_processes_RHI}
}
\footnotesize
\begin{tabular}{lp{6cm}}\tophline
\textbf{Checkpoint} & 
\textbf{Atmospheric processes} \\\middlehline
PBCDIAG    & Wet removal and resuspension of aerosols\\
RAD        & Radiation\\
PACEND     & Rayleigh friction and gravity wave drag\\
DYNEND     & Resolved dynamics and tracer transport\\
DEEPCU     & Deep convection\\
CLDMAC[01--06]  & Turbulence and shallow convection, sub-steps 1--6\\
CLDMIC[01--06]  & Stratiform cloud microphysics, sub-steps 1--6\\
\bottomhline
\end{tabular}
\end{table}

\subsubsection{Simulation setup}

The focus QoI in this example is
the relative humidity with respect to ice (RHI),
which directly affects the formation of new ice crystals. 
In EAMv1, ice nucleation is calculated after the parameterization 
of turbulence, shallow convection, and large-scale condensation
represented by
CLUBB \citep[Cloud Layers Unified By Binormals,][]{Golaz_et_al:2002, Larson_et_al:2002,Larson_and_Golaz:2005_MWR, larson_2017_tech_doc}.
CLUBB, ice nucleation, droplet nucleation, and other stratiform 
cloud microphysical processes represented by the parameterization 
of  \citet{Gettelman_et_al:2015_MG2_part1} are collectively sub-cycled
six times per $\Delta t_{\rm CPLmain}$.
Therefore in the namelist setup shown in Table~\ref{tab:namelist_RHI},
a checkpoint is selected before each invocation of the ice nucleation
parameterization (\texttt{CLDMAC01},..., \texttt{CLDMAC06}) to identify
sources of high RHI.
Additional checkpoints are selected after each invocation of 
the stratiform cloud microphysics 
(\texttt{CLDMIC01},..., \texttt{CLDMIC06}) 
to monitor how RHI decreases due to those processes.
A few other checkpoints are also selected to evaluate the impact of 
atmospheric processes that are known to affect air temperature
and specific humidity, for example large-scale dynamics, 
radiation, and deep convection.

In addition to monitoring RHI, we include
the specific humidity (Q) and the saturation specific 
humidity respect to ice (QSATI) as QoIs to 
help attribute the diagnosed RHI changes
(cf. namelist variable \texttt{qoi\_name} in Table~\ref{tab:namelist_RHI}). 
While Q is one of the prognostic variables in EAMv1,
RHI and QSATI need to be diagnosed at each checkpoint 
using three components of the model's prognostic state: 
Q, air temperature, and pressure. The diagnostic subroutines
are included in the module \texttt{misc\_diagnostics}.

All of the selected QoIs are
3D variables defined in 72 layers in EAMv1. 
Unlike in the previous example, \texttt{qoi\_x\_dp} is not specified
here; it gets the default values of zeros,
therefore no vertical integrals are calculated for the QoIs.

Two sampling conditions are specified: the first one selects
grid cells where RHI seen by the first invocation of the 
ice nucleation parameterization is higher than 125\%,
which is a necessary although insufficient condition to trigger 
homogeneous ice nucleation.
(For clarification, we note that 
RHI discussed here is the relative humidity calculated from the
grid-box mean specific humidity and grid-box mean air temperature. 
EAMv1 uses RHI~>~RH$_0$ as a screening
condition to determine if homogeneous 
ice nucleation can occur in a grid box. 
RH$_0$ depends on air temperature but has typical 
values around 125\%.)

The second condition effectively selects all grid cells and time steps, 
but we state the condition as RHI~$>$~-1\% instead of using the special 
metric ``ALL'', and select the same condition-evaluation checkpoint 
as in condition one, so that the conditionally sampled metric 
\texttt{cnd01\_RHI} and unconditionally sampled \texttt{cnd02\_RHI}
can be directly and conveniently compared.
(Using the special metric ``ALL'' would result in a metric variable 
\texttt{cnd02\_ALL}, which is a constant field of 1.0, 
being written to the output files.)

The checkpoint before the radiation
parameterization is considered 
the end of a full model time step and hence 
\texttt{cnd\_end\_chkpt} is set to \texttt{PBCDIAG}.
Both the field values and increments of the QoIs are monitored 
and included in model output.
The full set of fields tracked by our tool are sent to 
output tape 1 (the h0 file) which contains the monthly averages.

\begin{figure*}[t]
\includegraphics[width=17cm]{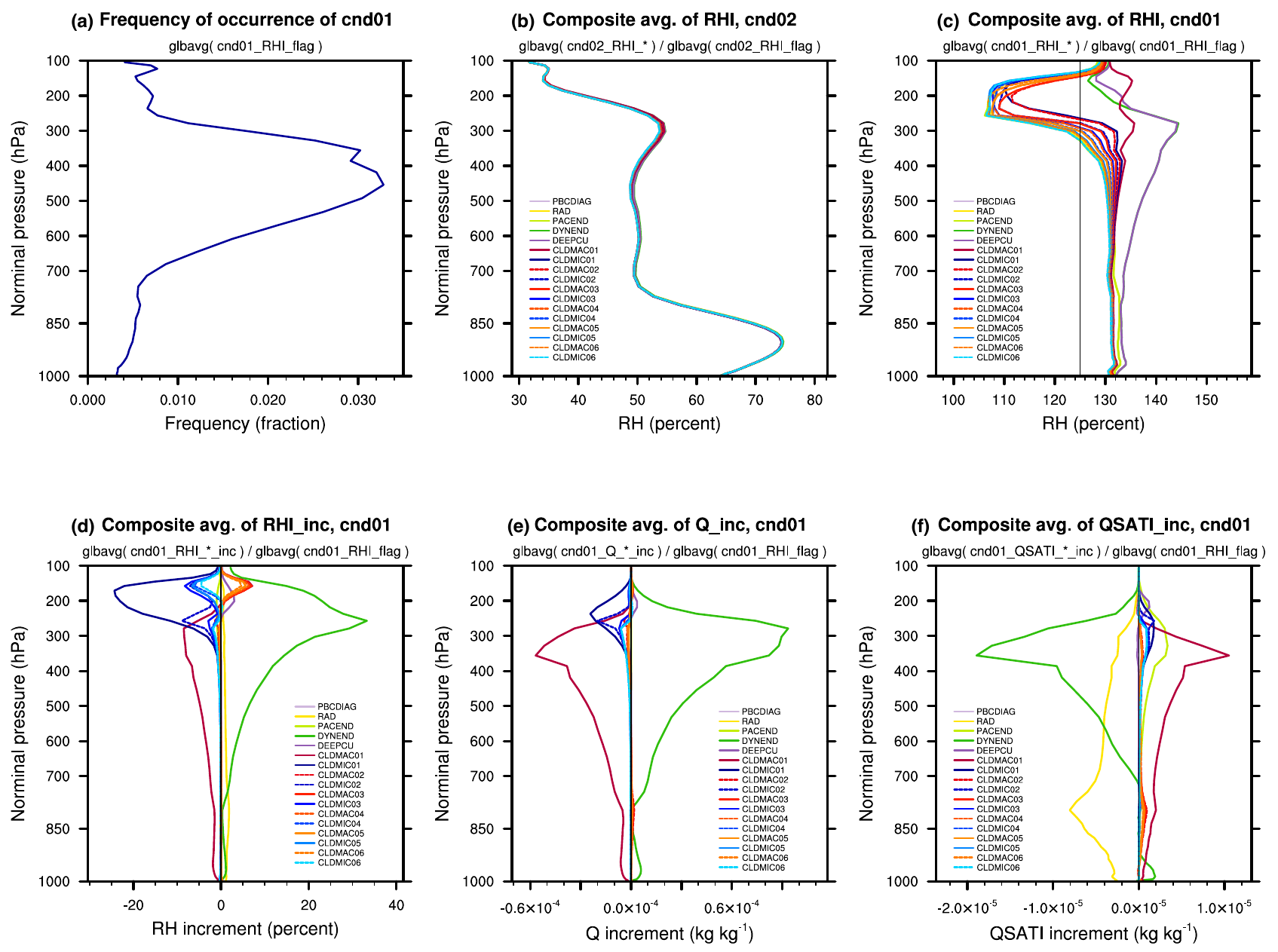}
\caption{
\label{fig:RHI_example_profiles}
Upper row:
(a) The frequency of occurrence of RHI~$>$~125\% 
averaged over one month and the entire globe.
(b) RHI at various checkpoints averaged over all time steps of the month and over the entire globe
(i.e., RHI under condition 02 -- unconditional sampling).
(c) Space-and-time mean RHI at various checkpoints under condition 01  
(i.e., RHI is higher than 125\% before the 
first ice nucleation calculation during a time step of $\Delta t_{\rm CPLmain}$~=~30~min).
Lower row: space-and-time mean increments of 
(d) RHI, (e) specific humidity,
and (f) saturation specific humidity with respect ice
averaged under condition 01.
}
\end{figure*}

\subsubsection{Results}

Figure~\ref{fig:RHI_example_profiles} shows various vertical profiles
derived from the simulation.
Defining a 2D global average 
as the average over all grid cells 
on a sphere weighted by their spherical area, panel (a) 
in Fig.~\ref{fig:RHI_example_profiles} shows the 
vertical profile of the 2D global average of the output variable
{cnd01\_RHI\_flag}, which gives the globally and temporally
averaged frequency of occurrence of RHI~>~125\% in each grid layer.
The other panels in the figure are global averages of 
different QoIs and checkpoints divided by the global mean frequency 
of occurrence of the corresponding condition.
Recall that our tool assigns a fill value of zero to 
grid cells and time steps that are unselected for a sampling condition.
The profiles in Figures~\ref{fig:RHI_example_profiles}b-f are therefore
spatial and temporal averages of the corresponding composites.

Panels (b) and (c) show RHI profiles under conditions 02 and 01, respectively.  
Sampling using the criterion of RHI~>~125\% 
helps to highlight the substantial changes related to ice cloud formation 
in the upper troposphere.
Panel (d) shows the increments of RHI at various checkpoints, 
allowing for a direct comparison of the signs and magnitudes 
of RHI changes caused by different physical processes.
The increments of specific humidity and saturation specific humidity
shown in panels (e) and (f) can further help to understand the 
physical mechanisms causing the RHI changes.


\conclusions[Conclusions and outlook]
\label{sec:conclusions}

An online diagnostic tool has been designed for and implemented in the 
global atmospheric circulation model EAMv1.
The motivation is to introduce a systematic way to support conditional sampling 
and budget analysis in EAM simulations, so as to 
(1) minimize the need for tedious ad hoc coding and hence save code development 
time and avoid clutter, and to
(2) reduce the need for instantaneous model output and hence
improve the computational efficiency of EAM simulations 
in which composite or budget analysis is needed.
 
Building upon the sequential process splitting used by EAM's time integration and 
the flexibility of the model's output functionalities,
the new tool adds its own data structures and functionalities to allow the users to 
select sampling conditions and model variables 
(also referred to as quantities of interest, QoIs) to monitor
at desired locations of the model's time integration cycles.
The condition metrics and QoIs can be any physical quantities
that are components of EAM's existing derived-type data structures
such as the physics state, physics buffer, and the data structures used 
for information exchanges between the atmosphere and the surface models 
such as land and ocean. The condition metrics and QoIs can also 
be any physical quantities that can be diagnosed from components 
of these existing data structures. 
Both the evolving values of the QoIs and their increments 
caused by different atmospheric processes can be monitored and 
written out as instantaneous or time-averaged values in EAM's output files
(also known as history tapes).
For QoIs defined at mid-points of the model's 
vertical grid or as layer averages, the tool also provides
the functionality to calculate and output vertical integrals 
weighted by the mass of dry or moist air.
Multiple sampling conditions can be used in a single simulation.
Unconditional sampling and  mixtures of conditional and unconditional sampling 
are also supported.

Assuming the user-chosen conditional metrics and QoIs as well as 
the locations in time integration cycle to monitor these quantities 
(referred to as checkpoints) 
are known to the tool, carrying out a composite or budget analysis 
using the new tool only requires setting a small number of namelist parameters.
The addition of new conditional metrics, QoIs, and checkpoints is straightforward
if the data to be sampled can be assessed through 
EAM's existing data structures.

The new tool has been designed for and implemented in 
EAMv1 and can be easily ported to 
EAMv1's descendants (e.g., EAMv2) or predecessors (e.g., CAM5) 
that use similar Fortran data structures
and time integration strategies.
Details of the design concepts  
and implementation in EAMv1 are explained in the paper
together with three use case examples that demonstrate 
the usage of the tool.

The development of the new tool was motivated by the need to carry out conditional budget 
analysis to understand sources of time-step sensitivities and time-stepping errors
related to EAMv1's physics parameterizations. 
While the current version of the tool, CondiDiag1.0, fulfills the authors' 
initial needs in those investigations,
we are aware of several aspects
in which the tool can be further extended or improved to benefit a wider range of EAM users:

First, if the desired condition metric or QoI is calculated 
by a lower-level (in software sense) subroutine and is not saved in 
EAM's derived-type data structures (e.g., physics state,
physics buffer, etc.), the most convenient way to pass 
data to CondiDiag would be to add the desired physical quantity to the physics buffer. 
Such cases will be further assessed and alternative methods will be explored. 
It is worth noting, however, that the E3SM project has been developing 
a brand new code base for its version 4 release. 
The new code uses a single ``field manager'' for information exchange 
between the host model and any resolved or parameterized atmospheric processes. 
An implementation of our tool in that new code base would make use of 
(and benefit from) this new ``field manager''.

Second, the specification of a sampling condition in CondiDiag1.0
takes the form of a logical expression involving the comparison of
a single metric with a threshold value. Section~\ref{sec:use_case_U10}
demonstrated how the tool can be used for a univariate probability 
distribution analysis.  
It would be useful to further extend 
the tool to support sampling conditions involving multiple metrics
and a series of threshold values for each metric, 
and hence facilitating multivariate probability distribution analysis. 
Along that line,
it might be useful to support sampling conditions involving 
multiple metrics evaluated at different checkpoints. 
This would be useful 
for investigating forcing-response relationships of multiple atmospheric processes
and for evaluating the behavior of sub-stepped model components.

Third, for simulations that involve multiple sampling conditions,
the current tool monitors the same set of QoIs and checkpoints 
under all conditions. While this will not be difficult to change,
we will assess the trade-off between more flexibility
and the potential risk of causing confusion for model developers and users.

Beyond the three aspects discussed above, there are some
desirable extensions of the tool that will require more substantial 
revisions of the current design. 
For example, in CondiDiag1.0, 
the sampling conditions are re-evaluated (and the QoIs are re-sampled) 
every model time step. We can, however, imagine cases where a user might 
want to evaluate a condition at some point of a simulation and 
monitor the evolution of the atmospheric state in the selected 
grid cells for longer time periods like a few hours or a few days. 
Supporting such use cases will require introducing an additional
mechanism to specify for how long the evaluated sampling condition is valid. 
Furthermore, 
anticipating possible modifications to the sequential splitting of atmospheric processes in EAMv1, in particular possible future adoption of parallel splitting or hybrid methods, it will be useful to explore how the current design of CondiDiag can be extended to accommodate other process coupling methods.

\codeavailability{
The EAMv1 code, run scripts, and postprocessing scripts used in this paper
can be found on Zenodo at \url{https://zenodo.org/record/5530189}.
Two versions of the EAM maint-1.0 code are provided:
one with CondiDiag1.0 implemented and one without.
} 

\clearpage
\appendix

\section{Candidate metrics and QoIs in CondiDiag1.0}
\label{apndx:available_metrics_and_QoIs}

Tables~\ref{tab:available_metrics_and_QoIs}--\ref{tab:available_metrics_and_QoIs_user}
list the currently available 
physical quantities 
that can be used as metrics for conditional sampling 
or be monitored as QoIs.


\begin{table*}[htbp]
\caption{Candidate condition metrics and QoIs that 
are directly copied from EAM's derived-type data structures.
``<cnst\_name>'' refers to tracer names in EAM.
``SF<cnst\_name>'' refers to variables names of 
tracer surface fluxes in EAM.
\texttt{pver} and \texttt{pverp} are EAM's variable names
for the number of vertical layers and vertical interfaces, respectively.
In the standard EAMv1, \texttt{pver} is 72 and \texttt{pverp} is 73.
The rightmost column explains the Fortran derived-type variables
and their components from which a metric or QoI's values
are obtained.
More candidate metrics and QoIs can be added following the example shown 
by the first code snippet
in Sect.~\ref{sec:key_algorithm_modules}.
\label{tab:available_metrics_and_QoIs}
}
\footnotesize
\begin{tabular}{|l|p{7cm}|l|l|}
\hline
\textbf{Name} & \textbf{Explanation}  & 
\textbf{Vertical dimension size} & \textbf{Data source} \\
\hline
<cnst\_name> & Advected tracers & \texttt{pver} & \texttt{state\%q} \\\hline
T & Air temperature & \texttt{pver} & \texttt{state\%t} \\\hline
U & Zonal wind & \texttt{pver} & \texttt{state\%u} \\\hline
V & Meridional wind & \texttt{pver} & \texttt{state\%v} \\\hline
OMEGA & Vertical velocity & \texttt{pver} & \texttt{state\%omega} \\\hline
PMID & Pressure at layer midpoints & \texttt{pver} & \texttt{state\%pmid} \\\hline
PINT & Pressure at layer interfaces & \texttt{pverp} & \texttt{state\%pint} \\\hline
ZM & Geopotential height at layer midpoints & \texttt{pver} & \texttt{state\%zm} \\\hline
ZI & Geopotential height at layer interfaces & \texttt{pverp} & \texttt{state\%zi} \\\hline
PS & Surface pressure  & 1 & \texttt{state\%ps} \\\hline
SF<cnst\_name> & Sfc. flux of advected tracers & 1 & \texttt{cam\_in\%cflx} \\\hline
LWUP & Longwave upward radiative flux from the surface & 1 & \texttt{cam\_in\%lwup} \\\hline
LHF & Latent heat flux from the surface & 1 & \texttt{cam\_in\%lhf} \\\hline
SHF & Sensible heat flux from the surface & 1 & \texttt{cam\_in\%shf} \\\hline
WSX & Surface stress (zonal) & 1 & \texttt{cam\_in\%wsx} \\\hline
WSY & Surface stress (meridional) & 1 & \texttt{cam\_in\%wsy} \\\hline
TREF & Ref. height surface air temp & 1 & \texttt{cam\_in\%tref} \\\hline
QREF & Ref. height specific humidity & 1 & \texttt{cam\_in\%qref} \\\hline
U10 & 10-m wind speed & 1 & \texttt{cam\_in\%u10} \\\hline
TS & Surface temperature & 1 & \texttt{cam\_in\%ts} \\\hline
SST & Sea surface temperature & 1 & \texttt{cam\_in\%sst} \\\hline
FLWDS & Downward longwave flux at surface & 1 & \texttt{cam\_out\%flwds} \\\hline
NETSW & Downward shortwave flux at surface & 1 & \texttt{cam\_out\%netsw} \\\hline
\end{tabular}
\end{table*}

\begin{table*}[htbp]
\caption{Candidate condition metrics and QoIs that 
are directly copied from EAM's ``physics buffer'' data structure.
\texttt{pver} and \texttt{pverp} are EAM's variable names
for the number of vertical layers and vertical interfaces, respectively.
In the standard v1 model, \texttt{pver} is 72 and \texttt{pverp} is 73.
More candidate metrics and QoIs can be added following existing
examples in subroutine \texttt{get\_values} in module \texttt{conditiona\_diag\_main}.
\label{tab:pbuf_metrics_and_QoIs}
}
\footnotesize
\begin{tabular}{|l|l|l|l|}
\hline
\textbf{Name} & \textbf{Explanation}  & 
\textbf{Vertical dimension size} & \textbf{Data source} \\
\hline
PBLH & Planetary boundary layer height & 1 & \texttt{pbuf} \\\hline
TKE  & Turbulent kinetic energy & \texttt{pverp} & \texttt{pbuf} \\\hline
UPWP & Turbulent momentum flux, east-west component  & \texttt{pverp} & \texttt{pbuf} \\\hline
VPWP & Turbulent momentum flux, north-south component& \texttt{pverp} & \texttt{pbuf} \\\hline
AST  & Stratiform cloud fraction & \texttt{pver} & \texttt{pbuf} \\\hline
CLD  & Total cloud fraction (stratiform plus convective) & \texttt{pver} & \texttt{pbuf} \\\hline
DEI  & Cloud microphysics: effective radius of cloud ice for radiation & \texttt{pver} & \texttt{pbuf} \\\hline
DES  & Cloud microphysics: effective radius of snow for radiation & \texttt{pver} & \texttt{pbuf} \\\hline
MU   & Cloud microphysics: size distribution shape parameter for radiation & \texttt{pver} & \texttt{pbuf} \\\hline
LAMBDAC  & Cloud microphysics: size distribution shape parameter for radiation & \texttt{pver} & \texttt{pbuf} \\\hline
\end{tabular}
\end{table*}

\begin{table*}[htbp]
\caption{Candidate condition metrics and QoIs 
that are diagnosed from components of EAM's derived type
data structures.
\texttt{pver} and \texttt{pverp} are EAM's variable names
for the number of vertical layers and vertical interfaces, respectively.
In the standard v1 model, \texttt{pver} is 72 and \texttt{pverp} is 73.
``Subroutine name'' is the name of the subroutine in module
\texttt{misc\_diagnostics} that calculates the requested quantity.
More candidates can be added following the 
the second code snippet in Sect.~\ref{sec:key_algorithm_modules}.
\label{tab:available_metrics_and_QoIs_user}}
\footnotesize
\begin{tabular}{|l|p{7cm}|l|l|}
\hline
\textbf{Name} & \textbf{Explanation}  & 
\textbf{Vertical dimension size} & \textbf{Subroutine name} \\
\hline
QSATW & Saturation specific humidity w.r.t. water & \texttt{pver} & \texttt{qsat\_water} \\\hline
QSATI & Saturation specific humidity w.r.t. ice & \texttt{pver} & \texttt{qsat\_ice} \\\hline
QSSATW & Supersaturation w.r.t. water  given as mixing ratio & \texttt{pver} & \texttt{supersat\_q\_water} \\\hline
QSSATI & Supersaturation w.r.t. ice  given as mixing ratio & \texttt{pver} & \texttt{supersat\_q\_ice} \\\hline
RHW & Relative humidity w.r.t. water in percent & \texttt{pver} & \texttt{relhum\_water\_percent} \\\hline
RHI & Relative humidity w.r.t. ice in percent & \texttt{pver} & \texttt{relhum\_ice\_percent}\\\hline
CAPE & Convective available potential energy & 1 & \texttt{compute\_cape}\\\hline
\end{tabular}
\end{table*}

\clearpage

\section{Checkpoints in CondiDiag1.0}    
\label{apndx:checkpoints}

Tables~\ref{tab:checkpoints_bc} and \ref{tab:checkpoints_ac} list
checkpoints currently implemented in EAM's physics driver subroutines
\texttt{tphysbc} and \texttt{tphysac}.
Table~\ref{tab:checkpoints_clubb_intr} lists
the current checkpoints in the interface subroutine \texttt{clubb\_tend\_cam}.


\begin{table}[htbp]
\caption{Checkpoints in the parameterization suite calculated before 
coupling with surface models, i.e., in the \texttt{tphysbc} subroutine.
\label{tab:checkpoints_bc}}
\footnotesize
\begin{tabular}{|l|l|ll}
\hline
\textbf{Model calculations after which} & \textbf{Checkpoint}  \\
\textbf{checkpoint is implemented}      & \textbf{name} \\\hline
Dynamical core  and large-scale transport & DYNEND \\\hline
Mass and energy fixers &  PBCINI \\\hline
Dry adiabatic adjustment &  DRYADJ \\\hline
Deep convection & DEEPCU \\\hline
Shallow convection (EAMv0 only) & SHCU \\     \hline                            
CARMA cloud microphysics & CARMA \\ \hline 
Stratiform cloud macrophysics, sub-step xx & CLDMACxx \\\hline   
Aerosol activation and mixing, sub-step xx & CLDAERxx \\\hline
Stratiform cloud microphysics, sub-step xx & CLDMICxx \\\hline
Stratiform clouds, all substeps & STCLD \\\hline
Aerosol wet removal and resuspension & AERWETRM \\ \hline    
Miscellaneous diagnostics and output & PBCDIAG \\   \hline 
Radiative transfer & RAD\\  \hline                     
Tropopause diagnosis; &\\
export state preparation and output  & PBCEND \\\hline
\end{tabular}
\end{table}

\begin{table}[htbp]
\caption{Checkpoints in the parameterization suite calculated 
after coupling with surface models, i.e., in the \texttt{tphysac} subroutine.
\label{tab:checkpoints_ac}}
\footnotesize
\begin{tabular}{|l|l|ll}
\hline
\textbf{Model calculations after which} & \textbf{Checkpoint}  \\
\textbf{checkpoint is implemented }     & \textbf{name}\\\hline
Couling to surface models & MCTCPL \\\hline
Emissions of chemical species & CHEMEMIS \\\hline
Tracer mass fixers & PACINI \\\hline
Chemistry and aerosol microphysics & CHEM\\\hline
Obukov length and friction velocity; &\\
Application of surface emissions & CFLXAPP \\\hline
Rayleigh friction & RAYLEIGH \\\hline
Aerosol dry deposition & AERDRYRM \\\hline
Gravity wave drag & GWDRAG \\\hline
QBO relaxation and ION drag & IONDRAG\\\hline
Application of nudging & NDG \\\hline
Dry-to-wet mixing ratio conversion & DRYWET \\\hline
Various diagnostics & PACEND \\\hline
\end{tabular}
\end{table}

\begin{table}[htbp]
\caption{Checkpoints implemented in the ``clubb\_tend\_cam'' 
subroutine.
\label{tab:checkpoints_clubb_intr}}
\footnotesize
\begin{tabular}{|l|l|ll}
\hline
\textbf{Model calculations after which} & \textbf{Checkpoint}  \\
\textbf{checkpoint is implemented}      & \textbf{name}\\\hline
Ice saturation adjustment, sub-step xx  & ICEMACxx \\\hline
CLUBB, sub-step xx  & CLUBBxx \\\hline
Convective detrainment, sub-step xx & CUDETxx \\\hline
Miscellaneous diagnostics, sub-step xx  & MACDIAGxx \\\hline
\end{tabular}
\end{table}

\noappendix       






\authorcontribution{
HW designed and implemented the diagnostic tool in EAMv1
with feedback from the coauthors.
KZ and HW designed the use case examples. 
HW carried out the simulations and processed the results.
HW wrote the manuscript; all coauthors helped with the revisions.
} 

\competinginterests{The authors declare no competing interests.} 


\begin{acknowledgements}
%
Computing resources for the initial development and testing of the new diagnostic tool were provided by the National Energy Research Scientific Computing Center (NERSC), a U.S. Department of Energy (DOE) Office of Science User Facility supported by the Office of Science of the U.S. Department of Energy under Contract No. DE-AC02-05CH11231.
Simulations shown as use case examples were carried out using the DOE Biological and Environmental Research (BER) Earth System Modeling program's Compy computing cluster located at Pacific Northwest National Laboratory (PNNL). PNNL is operated by Battelle Memorial Institute for the U.S. Department of Energy under Contract DE-AC05-76RL01830.
The EAMv1 code was obtained from the E3SM project sponsored by DOE BER
(DOI: 10.11578/E3SM/dc.20180418.36).\\

\noindent\textit{Financial support.}
This research has been supported by DOE BER 
via the Scientific Discovery through Advanced Computing (SciDAC) program
(grant no. 70276).
KZ was supported by DOE BER through the E3SM project (grant no. 65814).
\end{acknowledgements}

\end{document}